\documentclass[11pt]{article}%
\usepackage{algorithm}
\usepackage{algpseudocode}
\usepackage{amssymb}
\usepackage{subfigure}
\usepackage{amsfonts}
\usepackage{amsmath}
\usepackage{graphicx}%
\usepackage{multirow}
\providecommand{\U}[1]{\protect\rule{.1in}{.1in}}
\usepackage{tikz}
\usetikzlibrary{shapes}
\usepackage{pgfplots}
\usepackage{verbatim}
\usepgfplotslibrary{statistics}
\pgfplotsset{compat=1.8}
\usepackage{amsmath,amssymb,amsthm}
\usepackage{tikz}
\usetikzlibrary{positioning}

\usepackage{url}

\usepackage{xpatch}
\xpatchcmd\maketitle{\rlap}{}{}{}

\makeatletter
\newcounter{phase}[algorithm]
\newlength{\phaserulewidth}

\makeatother

\def\as{\textcolor{black}}

\def\asf{\textcolor{black}}

\usepackage{graphicx}

\begin{document}

\title{Network-based Prediction of COVID-19 Epidemic Spreading in Italy}
\author{Clara Pizzuti\thanks{National Research Council of Italy (CNR),
Institute for High Performance Computing and Networking (ICAR), Via P. Bucci, 8-9C, 87036, Rende, Italy; \emph{email}: clara.pizzuti@icar.cnr.it.}, Annalisa Socievole\footnotemark[1], Bastian Prasse\thanks{Faculty of Electrical Engineering, Mathematics and Computer Science, P.O Box 5031, 2600 GA Delft, The Netherlands.} and Piet Van Mieghem\footnotemark[2]}
%\date{Delft University of Technology \\
 %February 12, 2020}
\maketitle

\begin{abstract}
Initially emerged in the Chinese city Wuhan and subsequently spread almost worldwide causing a pandemic, the SARS-CoV-2 virus follows reasonably well the SIR (Susceptible-Infectious-\asf{Recovered}) epidemic model on contact networks in the Chinese case. In this paper, we investigate the prediction accuracy of the SIR model on networks also for Italy. Specifically, the Italian regions are \asf{a metapopulation} represented by network nodes and the network links are the interactions between those regions.  Then, we modify the network-based SIR model in order to take into account the different lockdown measures adopted by the Italian Government in the various phases of the spreading of the COVID-19. Our results indicate that the network-based model better predicts the daily cumulative infected individuals when time-varying lockdown protocols are incorporated in the classical SIR model.
\end{abstract}

\section*{Introduction}

The outbreak of the  greatest epidemic of the 21st century caused by the SARS-CoV-2 virus has stimulated researchers to understand and control
the spread of the disease inside a population with the help of mathematical models developed in recent years 
\cite{Hethcote00,PVM_RMP_epidemics2014}. 
A single outbreak of a disease is typically described by a SIR compartmental model, where each individual at a certain time $t$ can only be in one of the three different disease stages:   Susceptible (S), i.e. healthy, but vulnerable for the infection, Infected (I) and \asf{Recovered} (R), i.e. the individual either
recovers from the disease or, unfortunately, dies.  
A diffusion-like SIR epidemic spread on a contact network models the infection between individuals when they come into contact, close enough in space and long enough in time \cite{Chu_Lancet2020}. By adopting  the SIR model, Prasse et al. \cite{Prasse2020} predict the spreading of the COVID-19 
epidemic on a contact network consisting of 16 cities in the Chinese province Hubei via their Network Inference-based Prediction Algorithm (NIPA). Since the interactions between cities are unknown, Prasse et al. exploit their network reconstruction approach, described in \cite{PrasseM19}, to estimate
the contact network from the observations of the viral states.  

\begin{figure}
\centering
\includegraphics[width=0.4\linewidth]{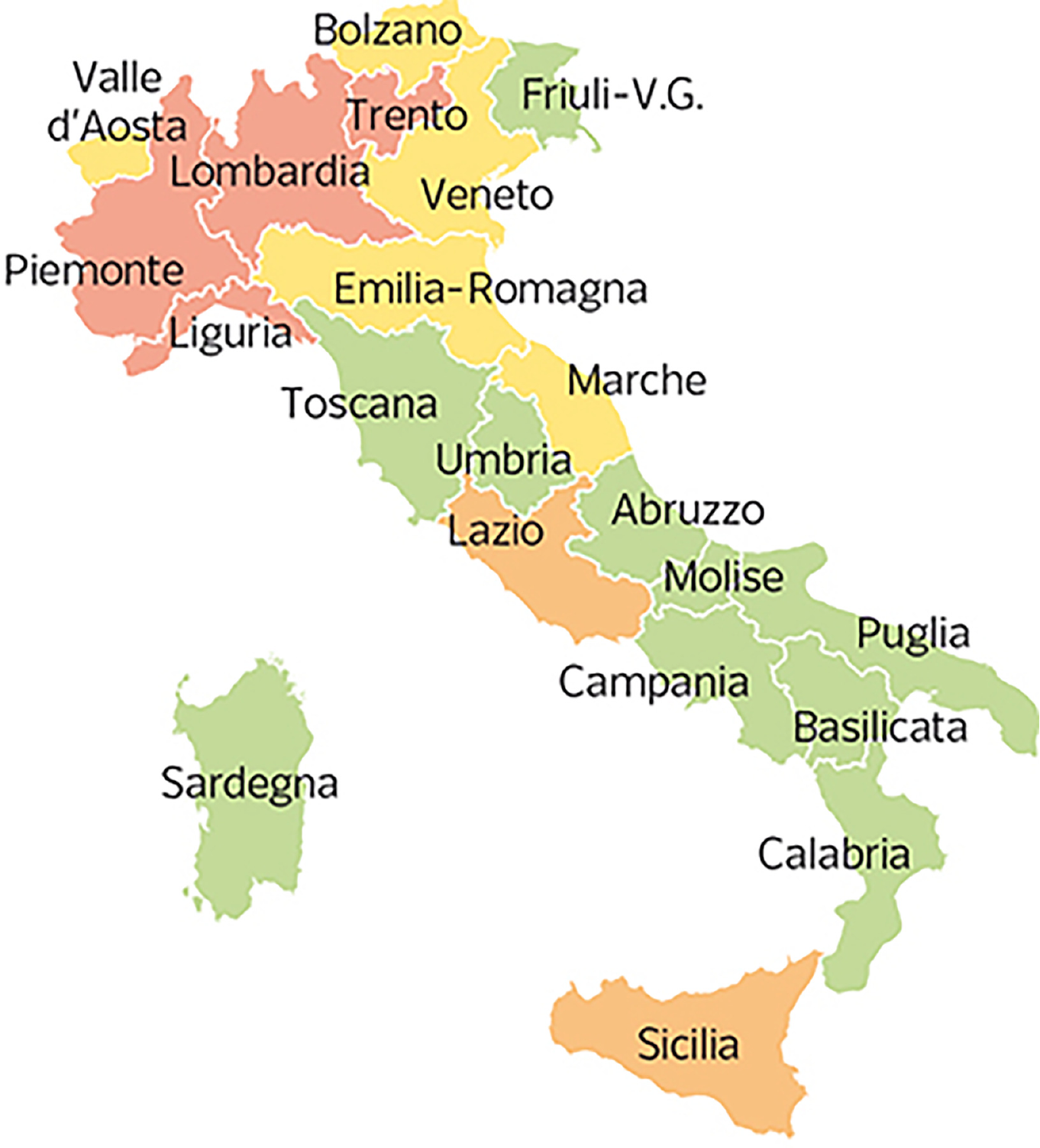}
\caption{The 21 Italian regions.}
 \label{fig:Italy_regions}
\end{figure}

In this paper, we use NIPA \cite{PrasseM19,Prasse2020} to investigate the spreading of the COVID-19 epidemic in Italy by considering the 21 Italian regions, shown in Fig. \ref{fig:Italy_regions}, as nodes of the network. We extend NIPA to NIPA-LD (NIPA with LockDown),  that takes into account the different {\em lockdown measures} adopted in the various phases of the COVID-19 spreading in Italy by adapting the ideas of Song et al. \cite{Wang2020}.
Song et al. \cite{Wang2020} pointed out that the  epidemiological models do not consider the several containment measures, such as in-home isolation, 
travel and social activities restrictions,  enforced by governments  to dampen the transmission rate over time. Due to the containment measures, the infection rates vary over time, which should be incorporated in a prediction  model to reflect the real situations of epidemic and provide more meaningful analyses.

We apply NIPA and the extension NIPA-LD to the period between the \asf{March 1 and June 9, 2020}. Our results indicate that NIPA-LD is capable to better predict the daily cumulative infected individuals, because the time-varying lockdown restrictions are considered.

%The paper is organized as follows. The next Section  describes some very recent approaches studying the effects of containment measures on the virus spread. Section \ref{sec:back} describes the $SIR$ model for COVID-19, which was considered for the original NIPA prediction method \cite{Prasse2020}. Section \ref{sec:esir} details the extended $SIR$ model with time-varying infection rates, which forms the basis for the extended prediction method NIPA-LD in this work. Section \ref{sec:tm} describes the transmission modifier adapted to the Italian case. Section \ref{sec:data}  introduces the regional data used for the COVID-19 spread in Italy. Section \ref{sec:tma} describes the analysis performed to choose a common transmission modifier for all the regions. Section \ref{sec:res} describes the experiments performed and the results obtained for assessing the accuracy of the model.  Finally, Section \ref{sec:concl} concludes the paper. 

\section*{Related work} \label{sec:rel}

In the last months, the number of papers studying the COVID-19 pandemic and proposing models to predict the evolution of the disease sky-rocketed. In \cite{estrada2020covid}, Estrada discusses how this pandemic is actually modeled and proposes future research directions by reviewing the three main areas of modeling research against COVID-19: epidemiology, drug repurposing, and vaccine design. After the strict policies in China to reduce close contacts between people, which revealed 
the best strategy to effectively block the virus diffusion,  Italy and many other European countries  imposed several containment measures, called lockdown. Some researches then investigated how mobility changed during the lockdown phases \cite{Oliver2020,Klein2020,Galeazzi2020,Schlosser2020}, others have shown how lockdown can effectively slow down disease transmission. Flaxman et al. \cite{flaxman2020estimating} study the effect on COVID-19 transmission of the major non-pharmaceutical interventions (NPIs) across 11 European countries for the period from the start of the COVID-19 epidemics in February 2020 until \asf{May 4} 2020. In a more general work, Haug et al. \cite{haug2020ranking} quantify the effectiveness of the world-wide NPIs to mitigate the spreading of COVID-19 and SARS-CoV-2 showing that this effectiveness is strongly related to the economic development as well as the dimension of governance of a country. At a country level, Hadjidemetriou at al. \cite{hadjidemetriou2020impact} use driving, walking and transit real-time data to investigate the impact of UK government control measures on human mobility reduction and consequent COVID-19 deaths. In \cite{pei2020differential}, Pei et al. assess the effect of NPIs on COVID-19 spread in the United States finding significant reductions of the basic reproductive numbers in major metropolitan areas when applying social distancing and other control measures. Di Domenico et al. \cite{di2020impact} study the case of the {\^I}le-de-France exploiting a stochastic age-structured transmission model which combines data on age profile and social contacts to evaluate the impact of lockdown and propose possible exit strategies. The Italian town of Vo' Euganeo is finally studied by Lavezzo et al. \cite{lavezzo2020suppression}, where the efficacy of the implemented control measures are evaluated, providing also insights into the transmission dynamics of asymptomatic individuals.

Concerning the modeling of the COVID-19 spreading with the imposed restrictions, Meyer and Brockmann \cite{MaierScience2020}, for instance,  proposed a model that takes into account both quarantine of symptomatic infected individuals and  population isolation due to containment policies, and showed that the model agrees with the observed growth of the epidemic in China. Arenas et al. \cite{Arenas2020} defined a model that stratifies the Spanish population by age and predicts the incidence of the epidemics through time by considering control measures. They show that the results can be refined by taking into account mobility restrictions imposed at the level of municipalities. Chinazzi et al. \cite{Chinazzi2020} used a global metapopulation disease transmission model to study the impact of travel limitations on the national and international spread of the epidemic in China. The NIPA-LD approach presented in this paper is different from the described proposals since it extends the NIPA method, which assumes no knowledge on the population flows and estimates the interactions between groups of individuals,  by considering time-varying lockdown policies in the prediction phase. 

Modeling the spread of COVID-19 in Italy has followed several approaches. Ferrari et al. \cite{ferrari2020modelling}, for instance, use an adjusted time-dependent SIRD (Susceptible-Infected-\asf{Recovered}-Died) model to predict the provincial cases.  In \cite{caccavo2020chinese}, Caccavo et al. propose a modified SIRD model to describe both the Chinese and the Italian outbreaks. Giuliani et al. \cite{giuliani2020modelling} define a model with $c = 8 $ compartments or stages of infection: susceptible (S), infected (I), diagnosed (D), ailing (A), recognized (R), threatened (T), healed (H) and extinct (E), collectively termed SIDARTHE. However, only one compartment is measured in the \asf{SARS-CoV-2} crisis, namely the number of active cases. Thus, for an epidemic model with many compartments, it is not possible to evaluate the accuracy in predicting compartments other than the number of active cases. In this work, we confine to the $c = 3$ compartmental SIR model for the predictions by NIPA. In \cite{kozyreff2020hospitalization}, Kozyreff provides \asf {a} SIR modeling comparison between Belgium, France, Italy, Switzerland and New York City suggesting that finer models are unnecessary with the corresponding available macroscopic data.

\section*{Background} \label{sec:back}
In this section, we briefly review the epidemic SIR  model on contact networks \cite{Scoglio2011,PrasseM19} and the prediction of the COVID-19 infection, caused by the  SARS-CoV-2 virus, based on the SIR model \cite{Prasse2020}. Then, we incorporate time-varying protocols introduced by the government to slow down the virus propagation.

We consider a network with $N$ nodes, where each node $i$ corresponds to the set of individuals living in the same place, like a city or a region.  An individual at any discrete time $k=1, 2, \ldots$ is in either one of the $c=3$ compartments \asf{Susceptible (S), Infectious (I), Recovered (R)}. The SIR model assumes that infectious individuals become \asf{recovered} and cannot infect any longer because of hospitalization, death, or quarantine measures. The \emph{viral state} of any node $i$ at time $k$ is denoted by the $3 \times 1$ vector $v_i[k] =(S_i[k],I_i[k],R_i[k])^T$, where $S_i[k],~I_i[k],~R_i[k]$ are the fractions of susceptible, infectious, and \asf{recovered} individuals, respectively, satisfying the conservation law $S_i[k]+I_i[k]+R_i[k]=1$.
The discrete-time SIR model \cite{Scoglio2011,PrasseM19} defines the evolution of the viral state $v_i[k]$ of each node $i$ as:
\begin{equation}\label{infe}
I_i[k+1] = (1 - \delta_i)I_i[k] + (1- I_i[k]  - R_i[k] )\sum\limits_{j=1}^N \beta_{ij} I_j[k] 
\end{equation}
\begin{equation}\label{rem}
R_i[k+1] = R_i[k] + \delta_i I_i[k] 
\end{equation}
where $\beta_{ij}$ denotes the \emph{infection probability} when individuals move from place (also called region) $j$ to place $i$. 
The self-infection probability $\beta_{ii}\neq 0$, because individuals inside the same place interact. The $N \times N$ infection probability matrix $B$ specifies the contact transmission chance between each couple of regions.   The \emph{curing probability} $\delta_i$ of place $i$ quantifies the capability of individuals in place $i$ to cure from the virus. 
We assume that the SIR model \asf{described in (\ref{infe}) and (\ref{rem})} has both $\beta_{ij}$ and $\delta_i$ that do not change over time. 

Prasse et al. \cite{Prasse2020} proposed the Network Inference-based Prediction Algorithm (NIPA), which estimates the spreading parameters  $\delta_i$ and $\beta_{ij}$ for each region $i$ from the time series $v_i[1], v_i[2], \ldots, v_i[n]$. These estimates in (\ref{infe}) and (\ref{rem}) predict the evolution of the virus in the next future times $k>n$.

\as{The SIR model has three compartments. In principle, with $c$ compartments, we must have $c-1$ independent measurements. 
The input to NIPA is only one  compartment, the infectious compartment $I$, which is less than \asf{$c-1=2$} compartments necessary  to reconstruct the network with the SIR model.  NIPA  creates observations of the R compartment by iterating over different candidate values of the curing rates $\delta_i$ and assuming the initial condition R(0)=0.
Thus, we observe only one compartment, the infectious compartment I, and the recovered compartment \asf{R}  is obtained by equation (2)  after estimating the curing probability $\delta_i$ in the training phase. }

To obtain the curing probability $\delta_i$, 50 equidistant values between $\delta_{min}$ and $\delta_{max}$ \asf{have been} considered, and then the value giving the best fit of model 
 (\ref{infe}) \asf{has been} used to estimate the matrix $B$ based on the least absolute shrinkage and selection operator (LASSO).
 \as{ 
For a general class of dynamics on networks (including the SIR model), completely different network topologies can result in the same dynamics. Hence, it is not possible to deduce the network accurately from observations, regardless of the reconstruction method: two very different networks perfectly match the observations, and there is no reason to infer  one network instead of the other. Thus, though NIPA accurately predicts the dynamics, the estimated network $B$ can be very different from the true network \cite{prasse2020predicting}. }

Let $n$ be  the number of days in which the infection has been observed. To evaluate the prediction accuracy, a fixed number of days $n_{neglect}$ is removed prior to $v_i[1], v_i[2], \ldots, v_i[n]$. The model is then trained on the days  $v_i[1], v_i[2], \ldots, v_i[n- n_{neglect}]$. Thereafter, the omitted $n_{neglect}$ days ($k=n- n_{neglect}+1, ..., n$) are predicted. It is possible to predict also $n_{predict}$ days ($k=n+1, ..., n+n_{predict}$) ahead the number $n$ of available observations, however, in such a case, we cannot evaluate the goodness of the prediction.

Prasse et al. \cite{Prasse2020} showed that the  approach accurately predicts the cumulative infections  for $n_{neglect} \leq 5$. However, if the number of neglected days increases, then the prediction capability of NIPA decreases. 
NIPA assumes constant values for $\beta_{ij}$, which, however, do not reflect the reality of the COVID-19 pandemic, because the containment measures imposed by the governments diminish $\beta_{ij}$ and thus the spread of the infection. 
Hence, infection probabilities $\beta_{ij}[k]$ \textit{which vary over time} $k$ should be considered in the epidemic model.

\section*{Extended SIR model with time-varying infection rate}\label{sec:esir}
Song et al. \cite{Wang2020} proposed the concept of \emph{transmission modifiers}, which decrease the  
probability that a susceptible individual can come into contact with an infected one because of the quarantine measures.

 At any discrete time $k$, let $q_S[k]$ be the chance of an individual to be in home isolation, and $q_I[k]$ the chance of an infected person to be in hospital quarantine. The \emph{transmission modifier} $\pi[k]$ is defined as follows:
 \begin{equation}
 \pi[k]=(1-q_S[k])(1-q_I[k]) \in [0,1]
 \end{equation}
and if no quarantine is active, then $\pi[k]=1$. 
In order to have a realistic infection rate $\beta$, Song et al. \cite{Wang2020}  multiply $\beta$ by 
$\pi[k]$ in the classic continuous SIR model.  Thus, the infection rate now reflects the effective currently enforced
quarantine measures taken in a country. In the extended SIR model, the curing probability $\delta_i$ remains the same, but the infection probability $\beta_{ij}$ is replaced by $\beta_{ij}\pi[k] $. The same considerations can be applied to the \textit{discrete-time} SIR model by modifying equation (\ref{infe}) above:

\begin{equation}\label{infem}
I_i[k+1] = (1 - \delta_i)I_i[k] + (1- I_i[k]  - R_i[k] )\sum\limits_{j=1}^N \beta_{ij} \pi[k]  I_j[k] 
\end{equation}

The \asf{transmission modifier} $\pi[k]$, however, should be specified on the base of the effective quarantine protocols 
undertaken in a specific region. Regarding the Hubei province in China, Song et al. \cite{Wang2020} suggest a step function mirroring the isolation measures established by the government. 

In the next section, the extended time-varying model (\ref{infem}) is applied to Italy by considering as nodes of the contact network the 21 regions by which Italy is composed. 
 
\section*{Transmission modifier for Italy}\label{sec:tm}
 
In Italy, the outbreak of the COVID-19 epidemic started in February in the North of Italy. A map of Italy with the division in regions is shown in Figure \ref{fig:Italy_regions}.  On February \asf{21}, the first case of infection appeared in the town of Codogno, in \textit{Lombardia}, and two cases also in the town of Vo' Euganeo in \textit{Veneto}. These two towns where immediately declared red zones and nobody could either go out or come in. On February \asf{24}, the three regions  of \textit{Lombardia}, \textit{Veneto}, and \textit{Emilia-Romagna} registered 172, 33, and 18 cases of infections, respectively. After that date, the virus propagated all over Italy  very fast.  

During the first week, until the first days of March,  no other particularly strict safety  measures were enforced. On March \asf{9}, however,   Italy turned into a lockdown \textit{Phase 1} with several strong restrictions and quarantine protocols. Schools, universities, shops, and many offices were closed, travels were not allowed and exits were only allowed for work, health or necessity situations with a mandatory self-certification. 

\textit{Phase 2} followed, in which countermeasures were adopted to reduce the pandemic. Finally, \textit{Phase 3} reopened almost all the activities and travels all over Italy. In order to define the values of the transmission modifier for the different quarantine periods,  we identified the following time intervals\footnote{Here, we recall the main reopening steps of commercial activities and services.}: 

\begin{itemize}
\item $\pi_0$: $k \le$ March 9  soft measures;
\item $\pi_1$: March 10 $\le k\le$ April 13 lockdown;
\item $\pi_2$:   April 14 $\le k \le$ May 3 libraries and stationeries reopen;
\item $\pi_3$:   May 4 $ \le k \le$ May 17 manufacturing, construction activities, wholesales reopen, meetings with relatives allowed; 
\item $\pi_4$:  May 18 $\le k \le$ May 24 hair dressers, beauty center, barber shops, bar, restaurants, retailers reopen, outdoor sport, baby parks allowed;
\item $\pi_5$:  May 25 $\le k \le$ June 2 gym, swimming pools, sport structures reopen
\item $\pi_6$:  $k \ge$ June 3 inter-regional mobility allowed.

\end{itemize}

The choice of the best values of the transmission modifier reflecting well the quarantine protocols is not an easy task and 
deserves a deep investigation. In the next sections, a study on the improvement of the NIPA method\asf{,} when different lockdown levels related to the  quarantine strategies \asf{are} adopted by authorities\asf{,} is performed.

\section*{Data preprocessing}\label{sec:data}
Our measurement data have been collected by the Italian Civil Protection Department\footnote{https://github.com/pcm-dpc/COVID-19} and are daily published on a repository. The available data are national, regional and provincial. We selected the regional ones which refer to the 21 regions depicted in Figure \ref{fig:Italy_regions}: \textit{Abruzzo}, \textit{Basilicata}, \textit{P.A. Bolzano}, \textit{Calabria}, \textit{Campania}, \textit{Emilia-Romagna}, \textit{Friuli Venezia Giulia}, \textit{Lazio}, \textit{Liguria}, \textit{Lombardia}, \textit{Marche}, \textit{Molise}, \textit{Piemonte}, \textit{Puglia}, \textit{Sardegna}, \textit{Sicilia}, \textit{Toscana}, \textit{P.A. Trento}, \textit{Umbria}, \textit{Valle d'Aosta}, \textit{Veneto}. 
Thus, for Italy, the entry
$\beta_{ij}$ of the $21\times 21$ matrix $B$ estimates the infection probability between the regions $j$ and $i$.
In the map, regions have been divided in 4 different colors representing the level of COVID-19 infected individuals. The \textit{red} regions 
have been the most affected by COVID-19, followed by the \textit{yellow} ones, the \textit{orange} ones and the \textit{green} regions with a lower number of cases.

For each observation day, we focused on the \textit{new positives} to COVID-19. 
We considered observations from \asf{March 1 to June 9}, 2020.

\section*{Transmission modifier analysis}\label{sec:tma}

To compare the NIPA method with the NIPA-LD implementing the lockdown measures, we 
considered  the model generated by NIPA which, in the training phase, neglects $n_{neglect}$ days, and then applied this model for the prediction phase by using different values of $\pi$
and an increasing  value of $n_{neglect}$. After that, we computed the average percentage error reduction  of NIPA-LD with respect to NIPA.  

Let $I_{CF,i}[k]$ be the observed cumulative fraction of infections of region $i$ at time $k$:

\begin{equation}
I_{CF,i}[k]=\sum_{\tau=1}^{k} I_{i}[\tau]
\end{equation}

To quantify the prediction accuracy we considered the \emph{Mean Absolute Percentage Error (MAPE)} defined as:
\begin{equation}
e[k]=\frac{1}{N}\sum_{i=1}^{N}{\frac{\mid I_{CFpred,i}[k]-I_{CF,i}[k]\mid}{I_{CF,i}[k]}}
\end{equation}
where $I_{CFpred,i}[k]$ is the predicted cumulative fraction of infected individuals in region $i$ at time $k$. 

Let $e[k]$ and $e_{LD}[k]$ denote the MAPE errors when $I_{CFpred,i}[k]$ is computed by NIPA and NIPA-LD, respectively. The percentage error improvement of NIPA-LD over NIPA is then computed as 

\begin{equation}
pe[k]=\frac{e[k] - e_{LD}[k]}{e[k]}\times 100
\end{equation}

In order to find a good transmission modifier which reflects the real situation best, we tested different  $\pi$ values by supposing a different response from people in respecting the quarantine measures imposed in the three months with varying levels of restrictions. Thus, we fixed increasing values of $\pi$ which intuitively correspond to a lower compliance to the containment protocols by the individuals. In view of the Italian lockdown measures previously described, we considered the following transmission modifier values:

$$\pi_{LD1}=[1~ 0.1 ~ 0.3~  0.5 ~ 0.7 ~ 0.8 ~ 1]  $$
$$\pi_{LD2}=[1~  0.2 ~ 0.4 ~ 0.6 ~ 0.8 ~ 0.9 ~ 1]  $$
$$\pi_{LD3}=[1~  0.3 ~ 0.5 ~ 0.7 ~ 0.85 ~ 0.95~  1]  $$
$$\pi_{LD4}=[1 ~ 0.4 ~  0.55 ~ 0.75 ~ 0.85 ~ 0.95 ~ 1]  $$
$$\pi_{LD5}=[1~  0.5 ~ 0.7 ~ 0.8 ~ 0.9~  0.95~  1]  $$
$$\pi_{LD6}=[1~ 0.6~  0.75~  0.85 ~ 0.95 ~ 0.99 ~ 1]  $$
$$\pi_{LD7}=[1 ~ 0.7 ~ 0.8 ~ 0.90 ~ 0.96~  0.99 ~  1]  $$

\begin{figure}	
\centering
\subfigure[]{
	   \includegraphics[width=0.45\linewidth]{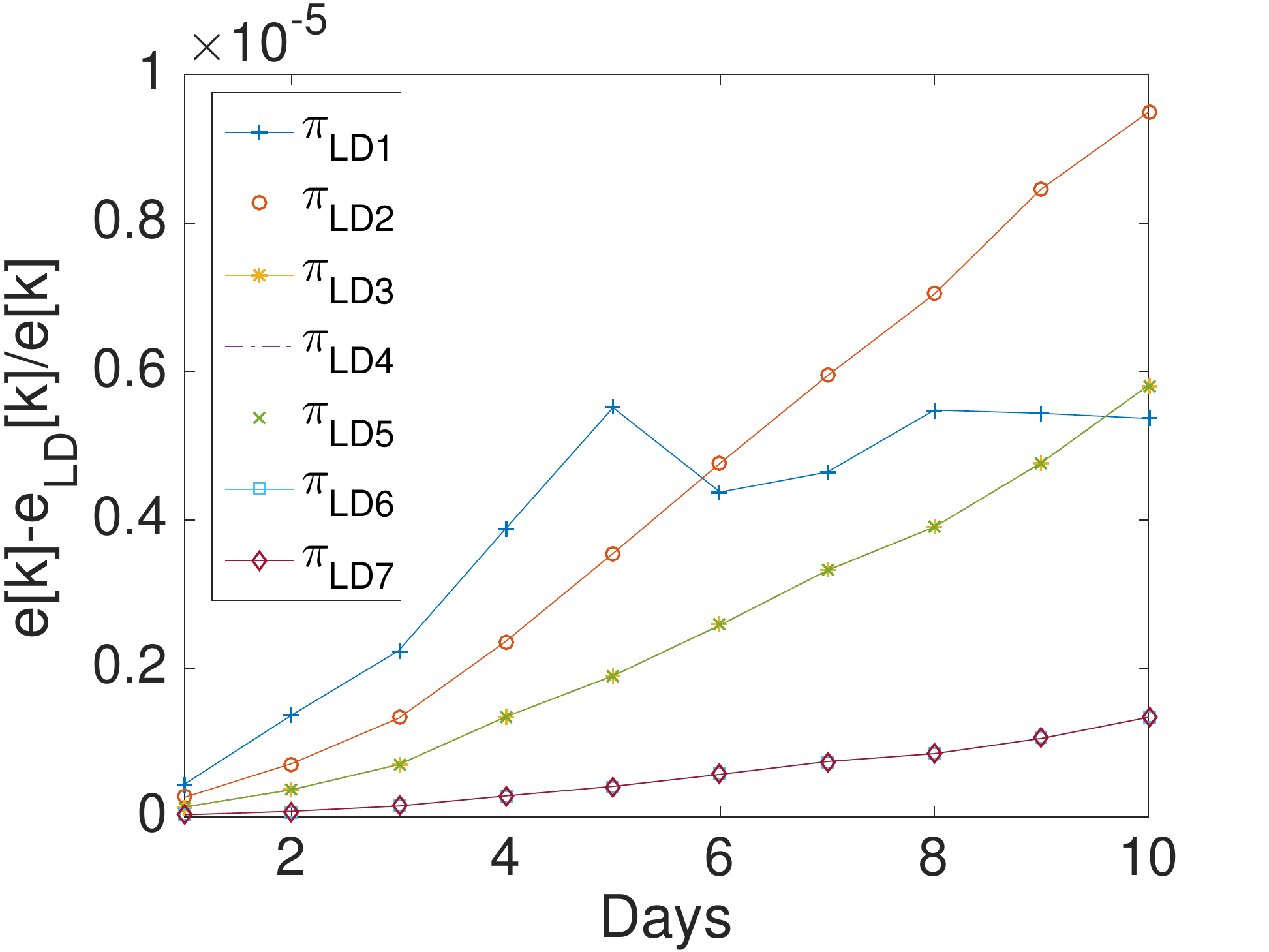}
  \label{fig:err1}}
\subfigure[]{
	   \includegraphics[width=0.45\linewidth]{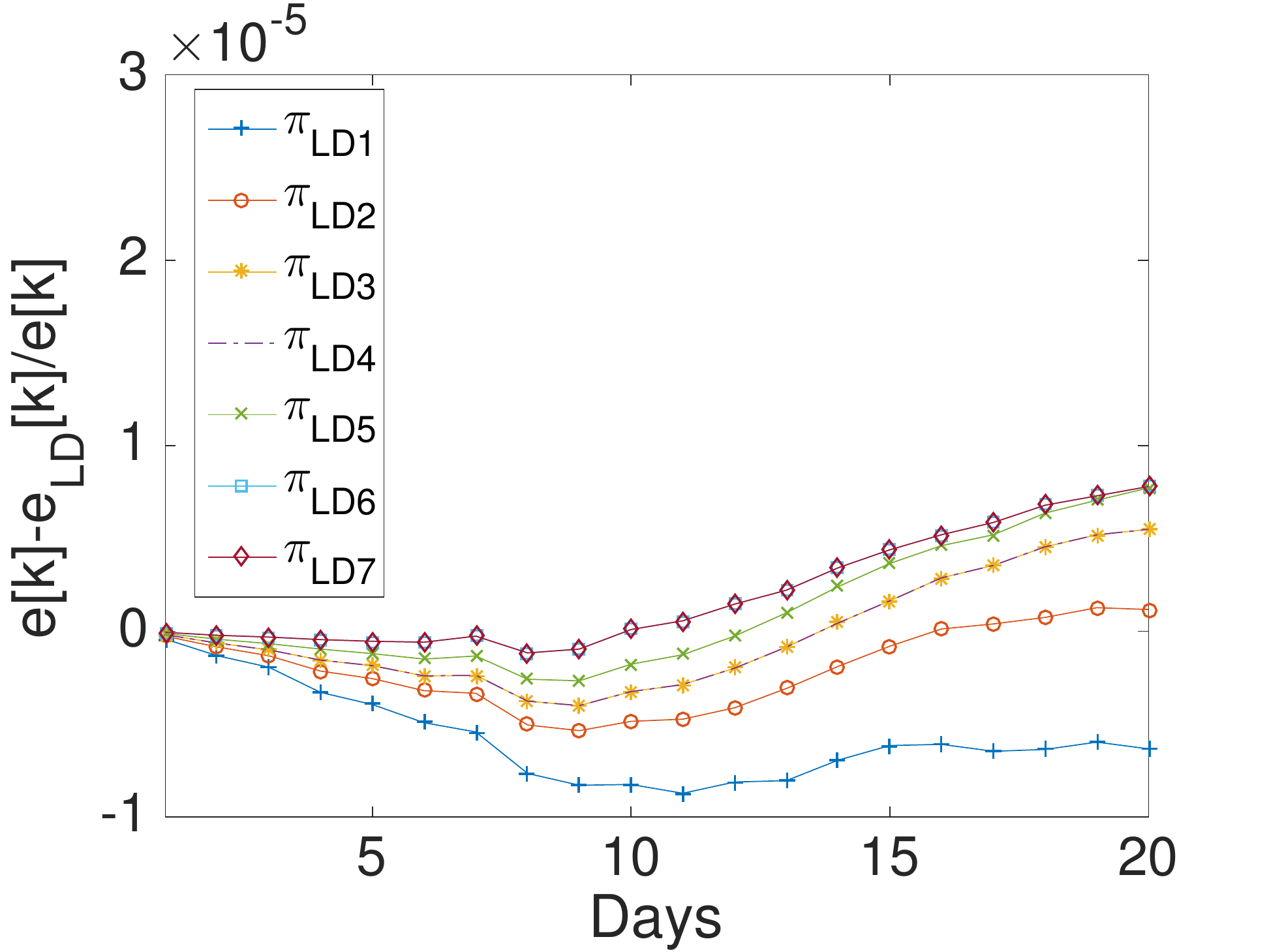}
	     \label{fig:err2}}
\subfigure[]{
	   \includegraphics[width=0.45\linewidth]{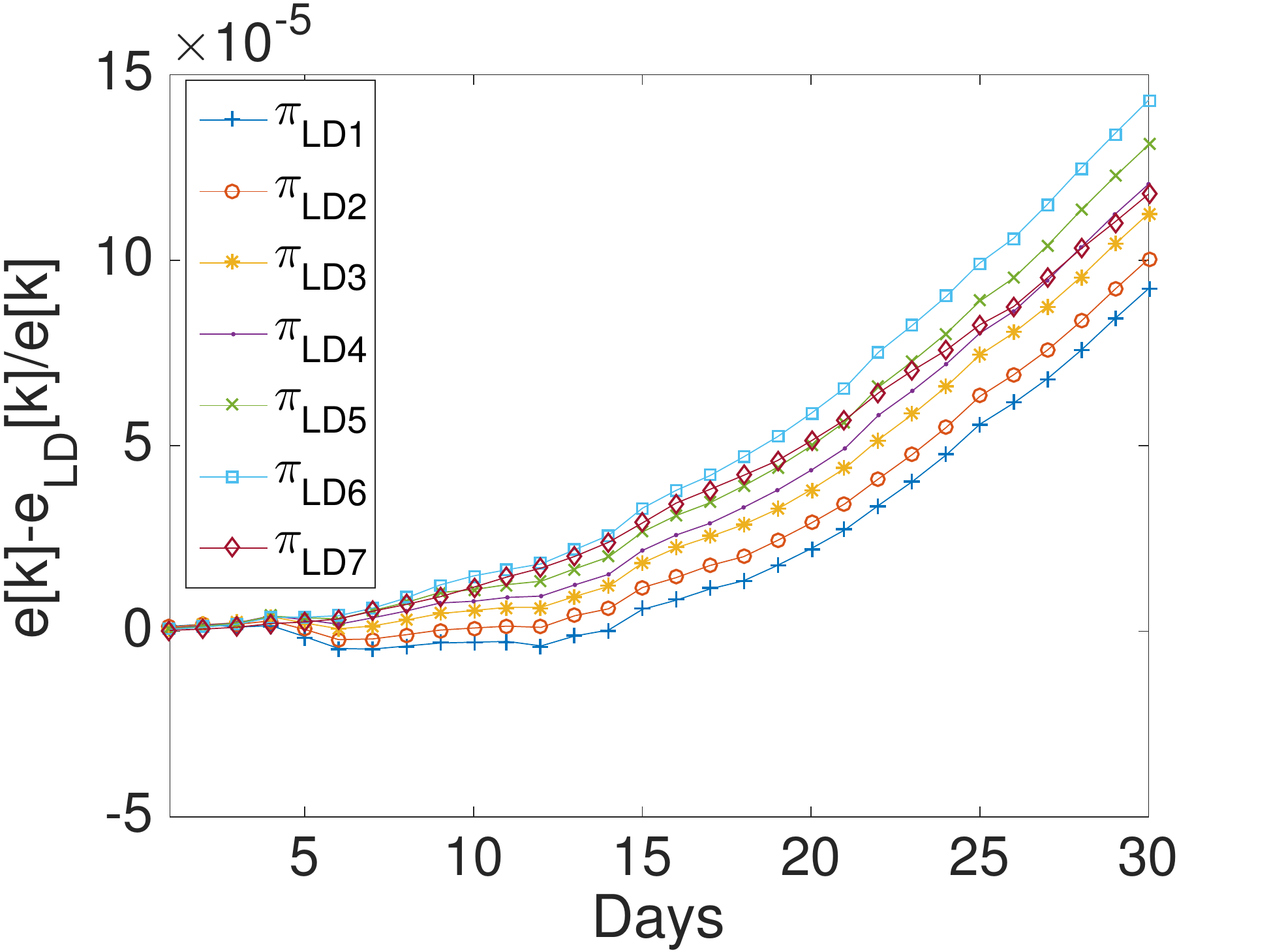}
	     \label{fig:err3}}
\subfigure[]{
	   \includegraphics[width=0.45\linewidth]{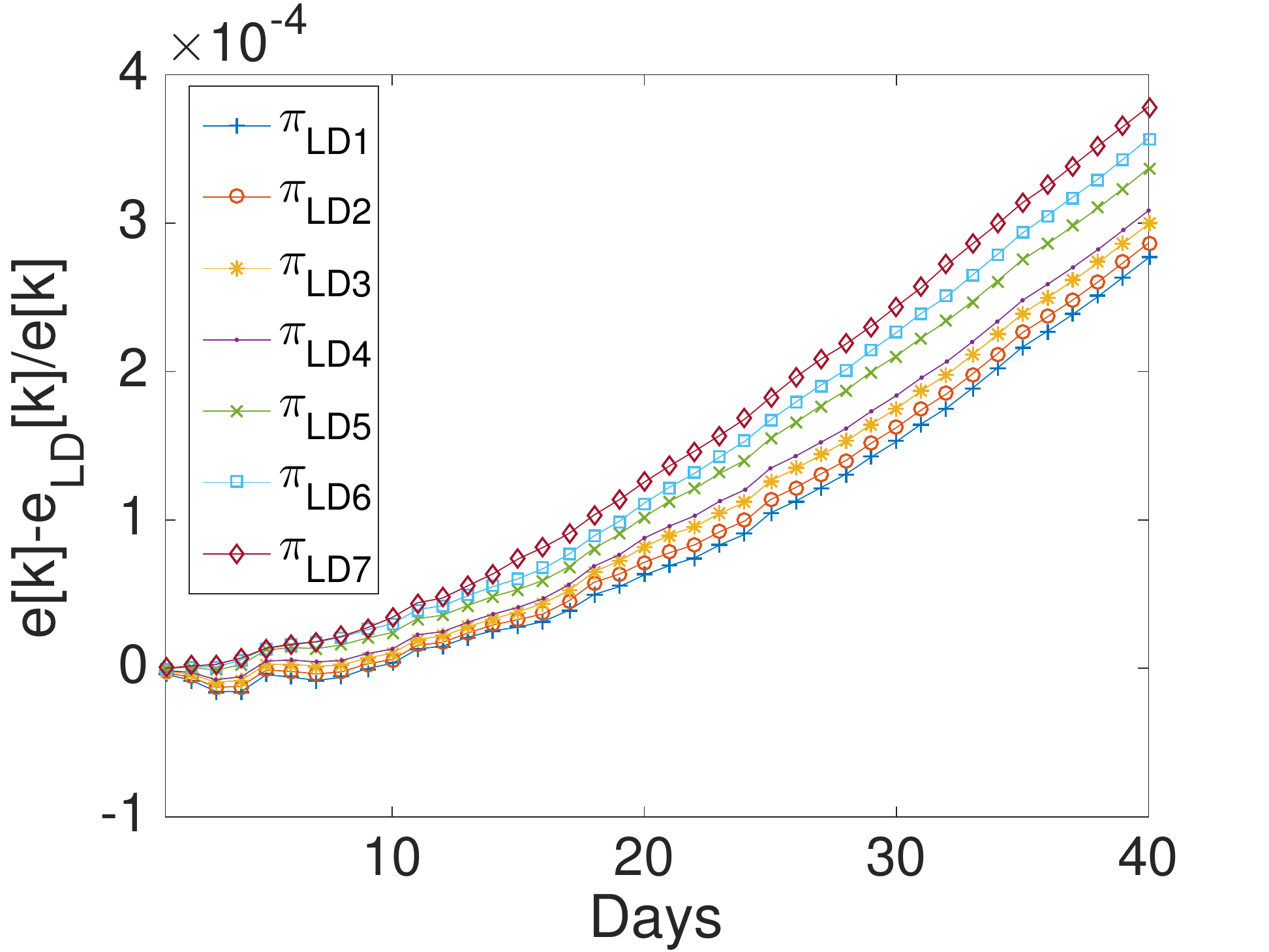}
	     \label{fig:err4}}
 \subfigure[]{
	   \includegraphics[width=0.60\linewidth]{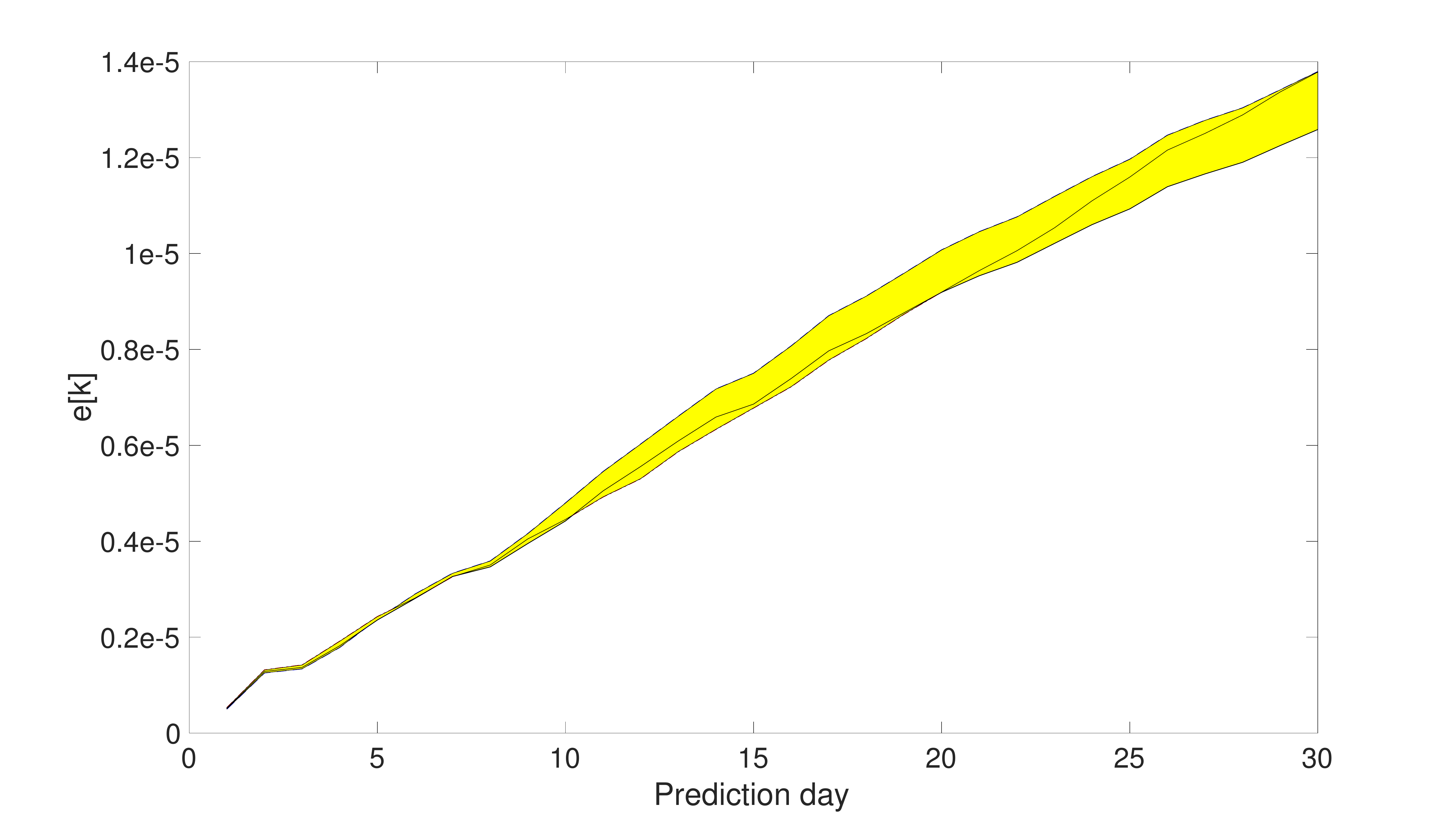}
	     \label{fig:cone}}
	     \caption{Mean prediction error when the number of the omitted days equals (a) $n_{neglect} =10$, (b) $n_{neglect}=20$, (c) $n_{neglect} =30$ and (d) $n_{neglect} =40$, for different transmission modifier vectors. (e) Cone of error evolution for $n_{neglect}=30$.}
\label{fig:err}
\end{figure}

Table \ref{tab:trial_error_p} reports the  improvement of the percentage error of NIPA-LD with respect to NIPA, for the  seven transmission modifiers and different numbers of predicted/omitted days, averaged over all the Italian regions and considering all the time windows under study, while Figure \ref{fig:err} shows the mean absolute prediction error as a function of the predicted/omitted days. From the table we can observe  that for $n_{neglect}$ equals to 10, 30 and 40 the percentage of improvement is overall very significant for most of the transmission modifier vectors. This means that NIPA-LD can be used to reliably perform both short and long term predictions. More specifically, for the short term predictions ($n_{neglect}=10$) low transmission modifier values are more suitable: $\pi_{LD1}$, for example, is able to achieve an improvement of 35.369\%. For the long term predictions, on the contrary,  where we neglect 30 or even 40 days aiming to predict them, higher transmission modifier values like those of $\pi_{LD7}$ perform better. When $n_{neglect} =20$ the error reduces, on average, only for  $\pi_{LD6}$ and  $\pi_{LD7}$. However, as Figure \ref{fig:err}(b) highlights, for  $\pi_{LD5}$  there is a reduction of the prediction error since the 10th day, and for $\pi_{LD4}$, $\pi_{LD3}$,  $\pi_{LD2}$ in the following next days, except for $\pi_{LD1}$.  Hence, for this case, we can conclude that soft lockdown protocols are able to induce a positive improvement in the error for all the values of  the number of neglected days. Finally, Figure \ref{fig:err}(e) depicts a cone of error evolution for $n_{neglect}=30$ when using  as transmission modifiers $\pi_{LD5},\pi_{LD6},\pi_{LD7}$, considering $\pi_{LD5}$ and $\pi_{LD7}$ as lower bound and upper bound of $\pi_{LD6}$, respectively. Then, we could assume that the future evolution of the epidemic can be predicted with an error that falls in between the predictions based on $\pi_{ub}$ and $\pi_{lb}$.

Figure  \ref{fig:err} shows that the differences between the different lockdown measures are meaningful\asf{. In} the next section, a detailed analysis  
for all the Italian regions  is performed to evaluate  the prediction accuracy of NIPA and NIPA-LD. 

\begin{table}
%\scriptsize
%\footnotesize
\centering
\caption{Percentage improvement of NIPA-LD over NIPA prediction for different transmission modifier values and increasing number of neglected days.}
\label{tab:trial_error_p}
\begin{tabular}{|c|c|c|c|c|c|c|c|} \hline
%\hline\noalign{\smallskip}
{\bf $n_{neglect}$} & {\bf $pe_{\pi_{LD1}}$ } & {\bf $pe_{\pi_{LD2}}$ } & {\bf $pe_{\pi_{LD3}}$ } & {\bf $pe_{\pi_{LD4}}$ } & {\bf $pe_{\pi_{LD5}}$ } & {\bf $pe_{\pi_{LD6}}$ } & {\bf $pe_{\pi_{LD7}}$ } \\ \hline
%\noalign{\smallskip}\hline\noalign{\smallskip}
10 & 35.369  &  34.566  &  19.279  &  19.279  &  19.279  &  4.22  &  4.22 \\ \hline
20 & -33.766  &  -16.233  &  -7.7 &  -7.7  &  -0.842  &  4.805  &  4.805  \\ \hline
30 &   10.438  & 15.147 &  20.894 &  23.981 &  27.721 &  31.747 &  26.56  \\ \hline
40 & 17.921 &  22.802 &  28.729 &  32.319 &  43.716 &  49.14  & 54.095  \\ \hline
\end{tabular}
\end{table}

\section*{Results} \label{sec:res}
In this section, we evaluate  the prediction accuracy of NIPA and NIPA-LD by  computing the cumulative infections for each observation day  when $n_{neglect}=30$ and  compare them to the true data by using $\pi_{LD6}$ as  transmission modifier for the different quarantine periods. In this experiment, thus, NIPA does not consider the 30 last days of the observed daily data of the newly infected individuals for estimating the curing probability $\delta_i$ and the infection probability $\beta_{ij}$. Then\asf{,} both NIPA and NIPA-LD  predict  the cumulative infections from May~10 until June~9 \asf{and the predictions} are compared with (a) the true data, and (b) to the logistic function as baseline. 
The logistic function, introduced in the 19th century by Verhulst to model population growth, approximates the solutions of the SIS and SIR models \cite{Kermack,prasse2020fundamental}.  
The cumulative number of infected cases $y_i[k]$ at time $k$ for the region $i$ is assumed to follow:
\begin{equation}
y_i[t]=\frac{y_{\infty,i}}{1+e^{-K_i(t-t_{0,i})}}
\end{equation}
where $y_{\infty,i}$ is the long-term fraction of infected individuals, $K_i$ is the logistic growth rate,  $t$ is the time in day.

\begin{figure}	
\centering
\includegraphics[width=0.7\linewidth]{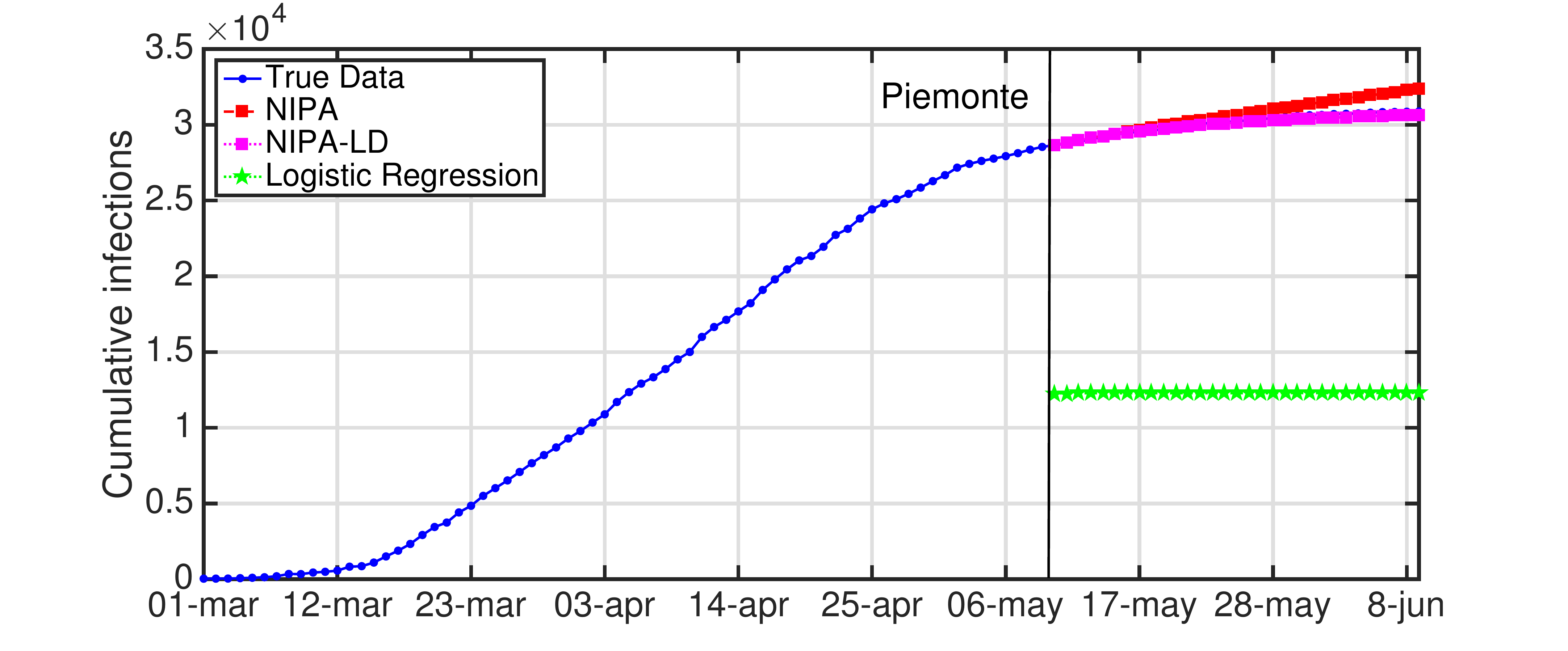}
\caption{Cumulative infections for \textit{Piemonte}.}
\label{fig:Piemonte}
\end{figure}

\begin{figure}	
\centering
\includegraphics[width=0.7\linewidth]{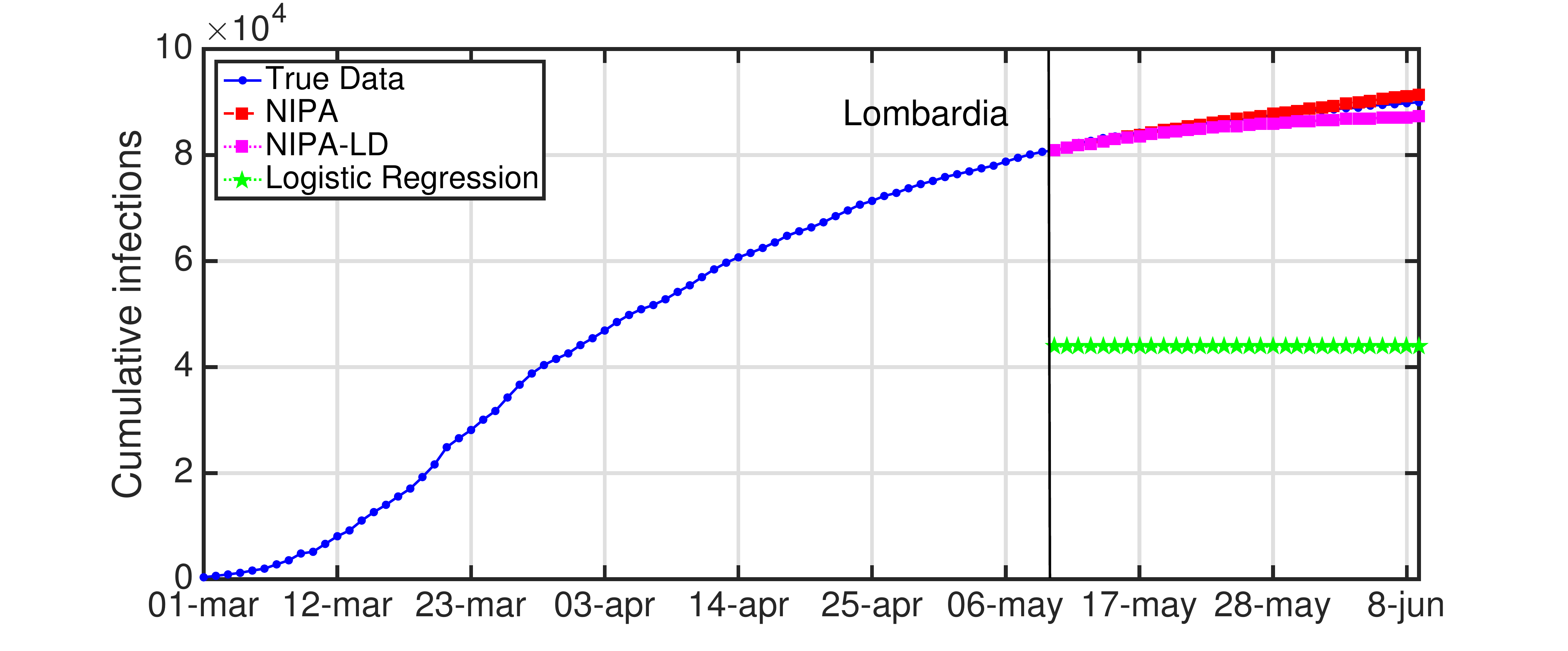}
\caption{Cumulative infections for \textit{Lombardia}.}
\label{fig:Lombardia}
\end{figure}

\begin{figure}	
\centering
\includegraphics[width=0.7\linewidth]{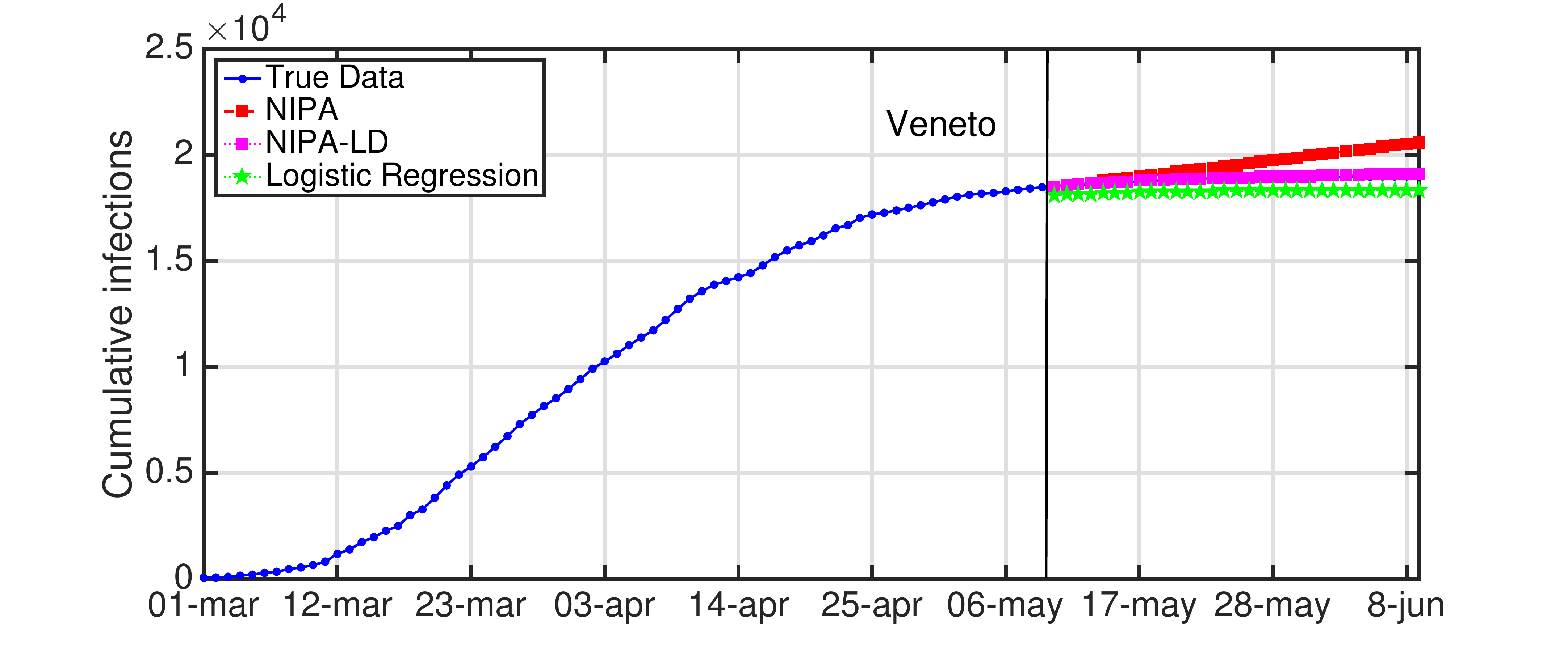}
\caption{Cumulative infections for \textit{Veneto}.}
\label{fig:Veneto}
\end{figure}

\begin{figure}	
\centering
\includegraphics[width=0.7\linewidth]{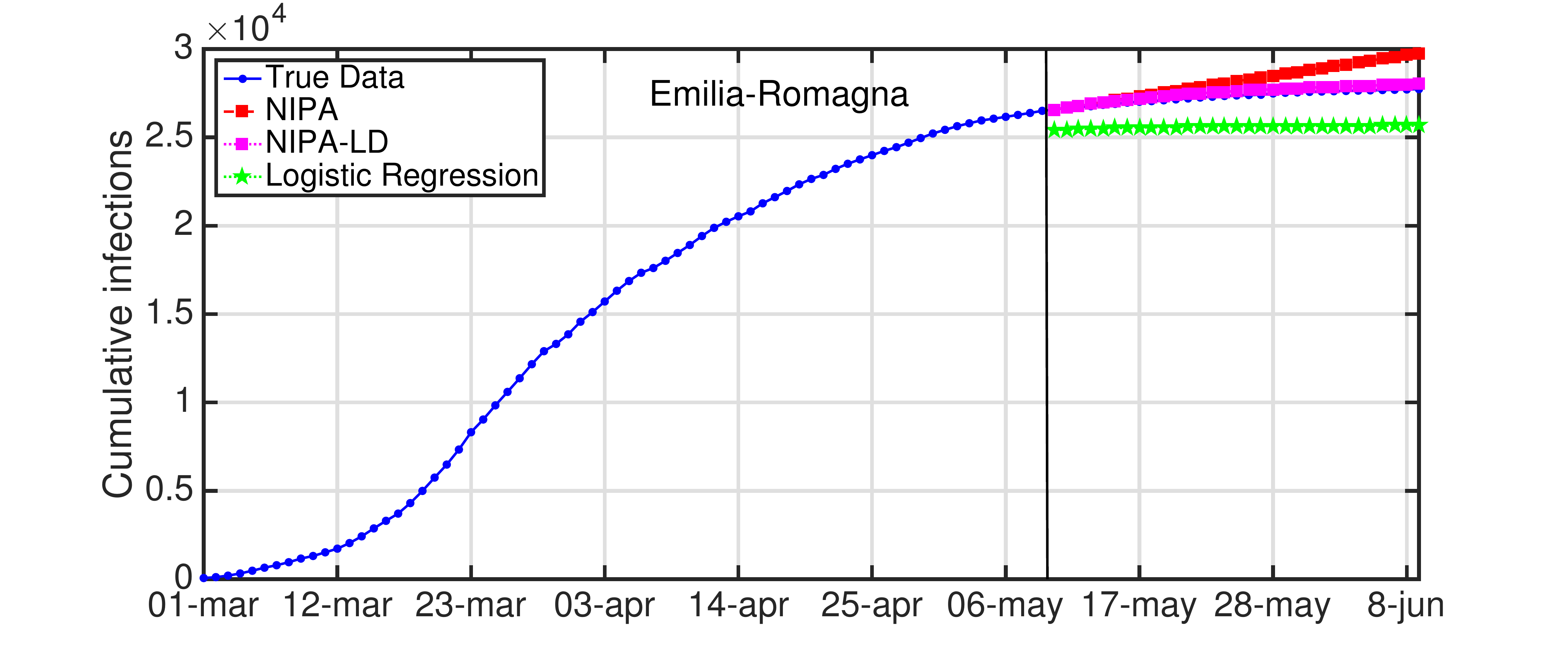}
\caption{Cumulative infections for \textit{Emilia-Romagna}.}
\label{fig:EmiliaRomagna}
\end{figure}

\begin{figure}	
\centering
\includegraphics[width=0.7\linewidth]{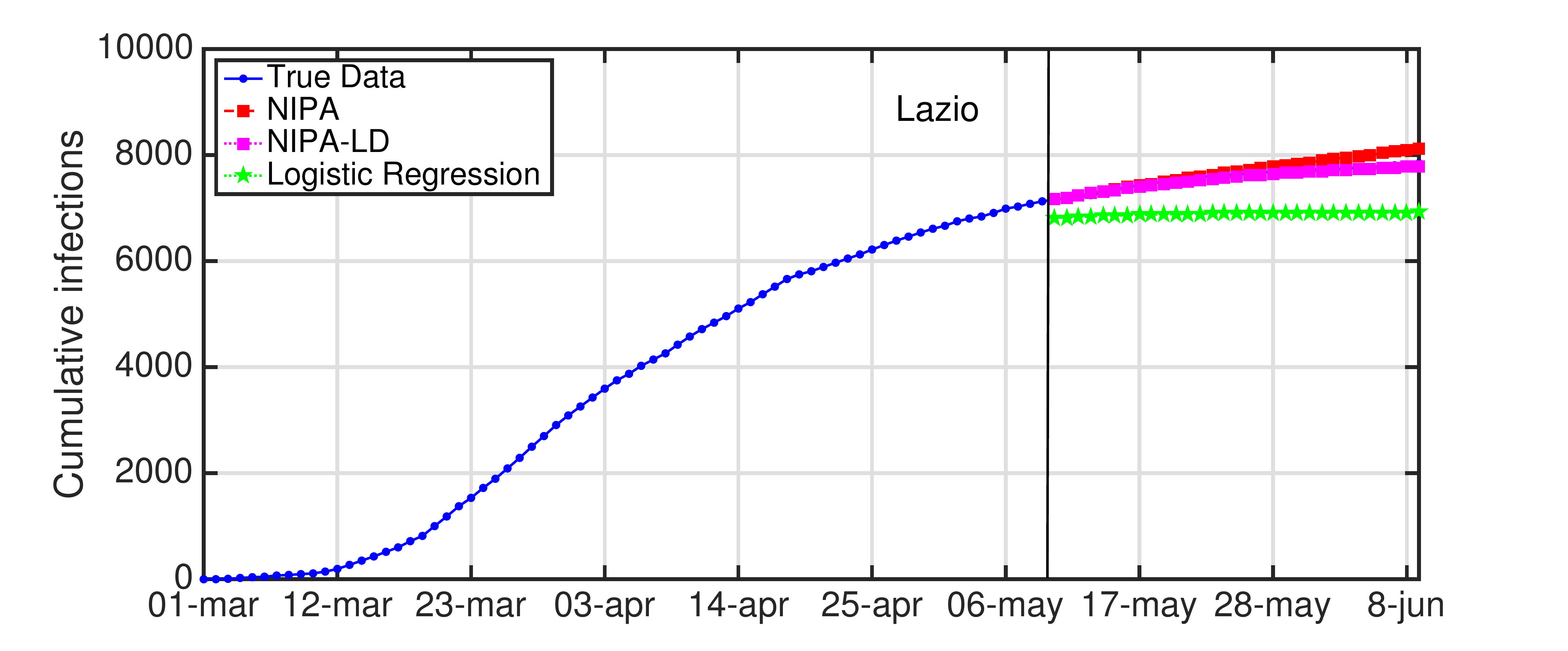}
\caption{Cumulative infections for \textit{Lazio}.}
\label{fig:Lazio}
\end{figure}

\begin{figure}	
\centering
\includegraphics[width=0.7\linewidth]{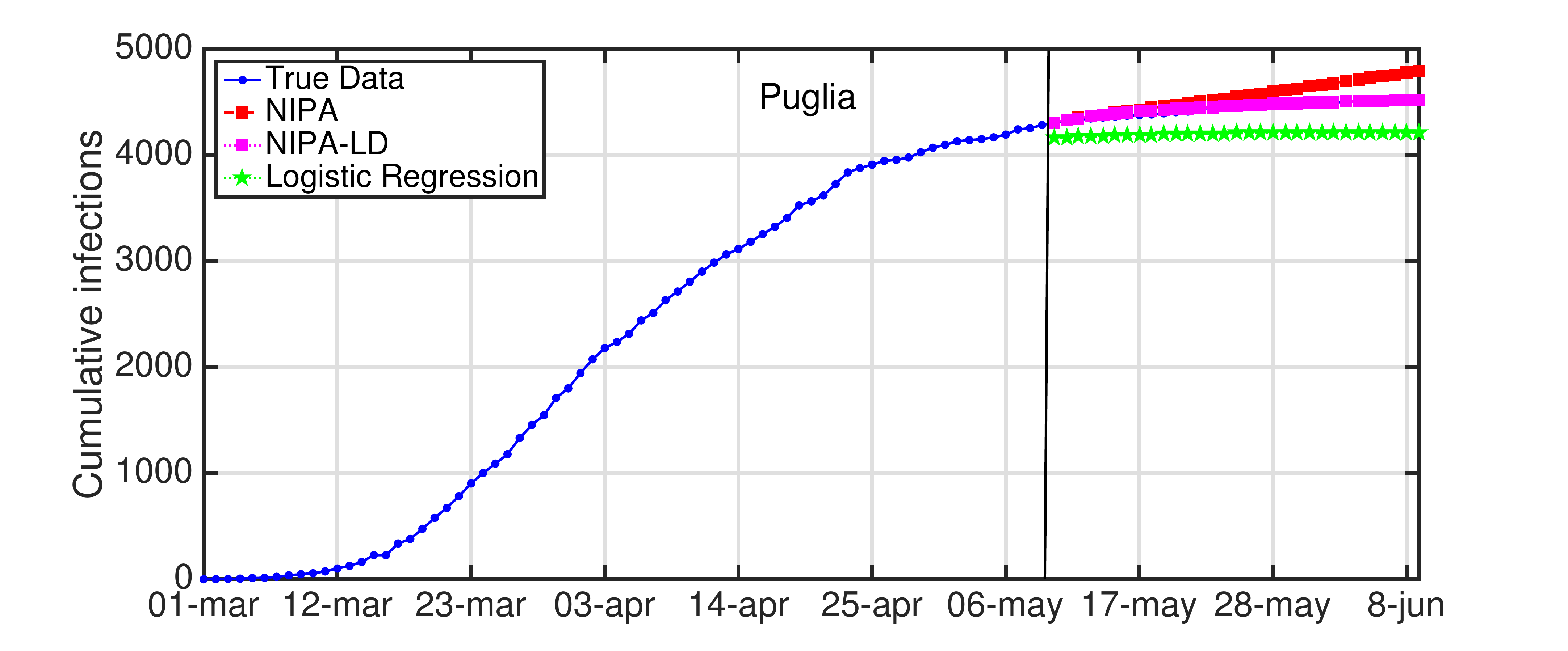}
\caption{Cumulative infections for \textit{Puglia}.}
\label{fig:Puglia}
\end{figure}

Due to lack of space, we only report the plots for a subset of the North regions, the ones highly affected by the virus spreading in the red and yellow zones (\textit{Piemonte}, \textit{Lombardia}, \textit{Veneto}, \textit{Emilia-Romagna}), for one representative region of the orange zone (\textit{Lazio}) and for one of the green zones (\textit{Puglia}). For the center and the south of Italy, the COVID-19 spreading has been characterized by a lower number of cases and for this reason we report only two representative regions. In Figure \ref{fig:Piemonte}, the cumulative infections for \textit{Piemonte} are shown. Here, the lockdown modified NIPA variant clearly outperforms the classical NIPA, which overestimates the number of infected individuals. For \textit{Piemonte}, NIPA-LD better matches the true data. Moreover, for this region, a simple logistic regression is not able to well predict the epidemic.  Figure \ref{fig:Lombardia} depicts the trend of the predictions for the most challenging region in Italy, \textit{Lombardia}, which has been mostly affected by the COVID-19. Again, the logistic regression excessively understimates the cumulative infections. From May 10 to May 30, both  NIPA and NIPA-LD  models well match the number of cumulative infections. However, for the next days, NIPA slightly overestimates the infections while NIPA-LD underestimates them. This is probably due \asf{to} a much higher mobility of the population after the loosening of the lockdown rules on May 25. 
The \textit{Veneto} case (Figure \ref{fig:Veneto}), another region of the North Italy highly affected by the COVID-19, on the contrary, is accurately predicted by NIPA-LD, while classical NIPA without lockdown clearly overestimates the number of infections. Here, the logistic regression works better than the previous regions but still understimates the cumulative infections. For the last North region, \textit{Emilia-Romagna}, the cumulative infections are better predicted by the lockdown modified NIPA, which slightly overestimates the infections but to a lesser extent than the classical NIPA (Figure~\ref{fig:EmiliaRomagna}). The baseline on the contrary, underestimates the infections. In Figure \ref{fig:Lazio}, the results for \textit{Lazio} confirm the better accuracy of NIPA-LD. Finally, Figure \ref{fig:Puglia} shows the results obtained for the \textit{Puglia} region. We observe that the NIPA prediction with the lockdown transmission modifiers is able again for this region to accurately predict the cumulative infections, while the classical NIPA overestimates them from May 15 until June 9 and the logistic regression underestimates the infections even from May 10.

\begin{figure}	
\centering
\subfigure[]{
	   \includegraphics[width=0.45\linewidth]{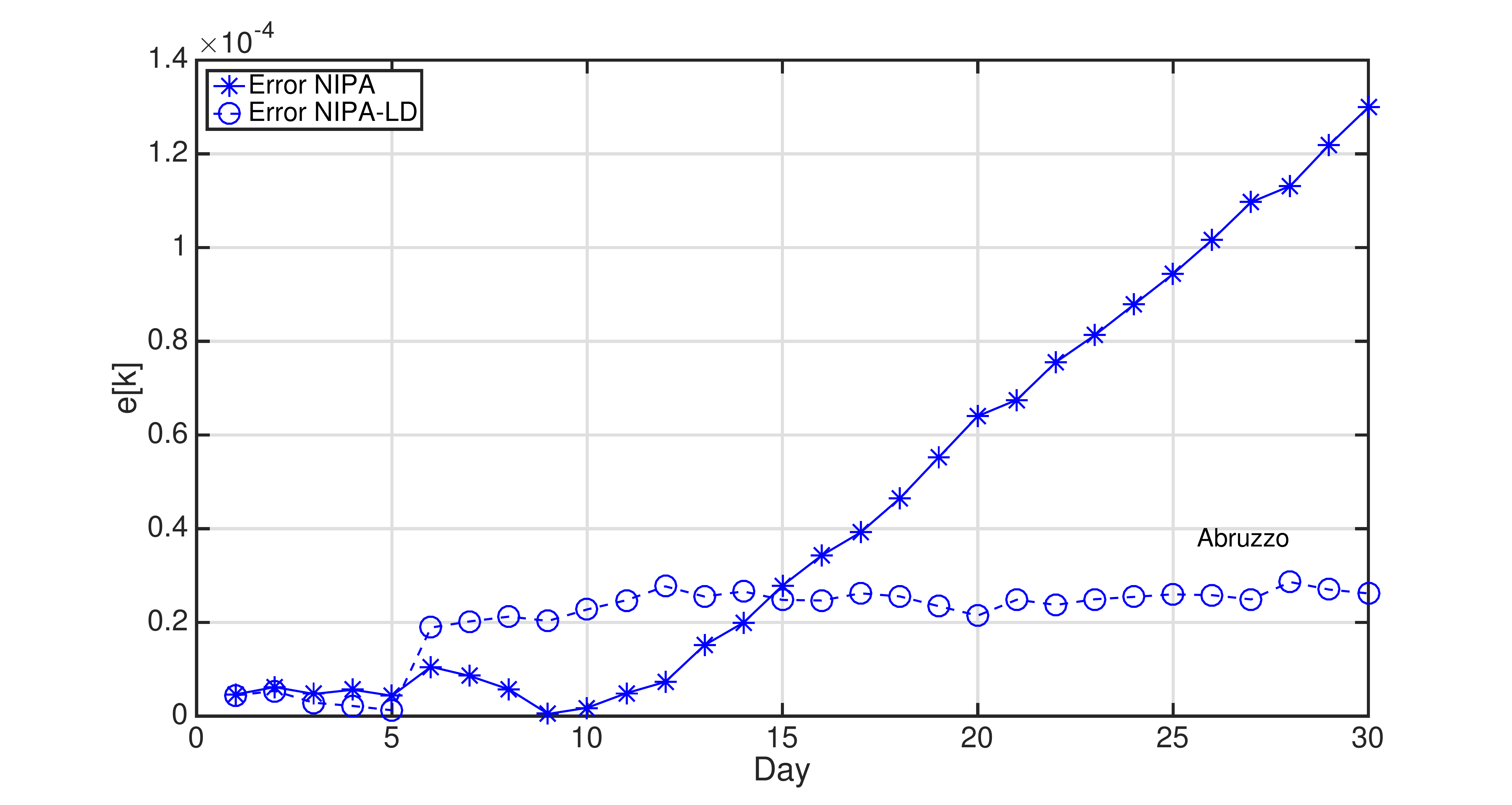}
	     \label{fig:Ab}}
\subfigure[]{
	   \includegraphics[width=0.45\linewidth]{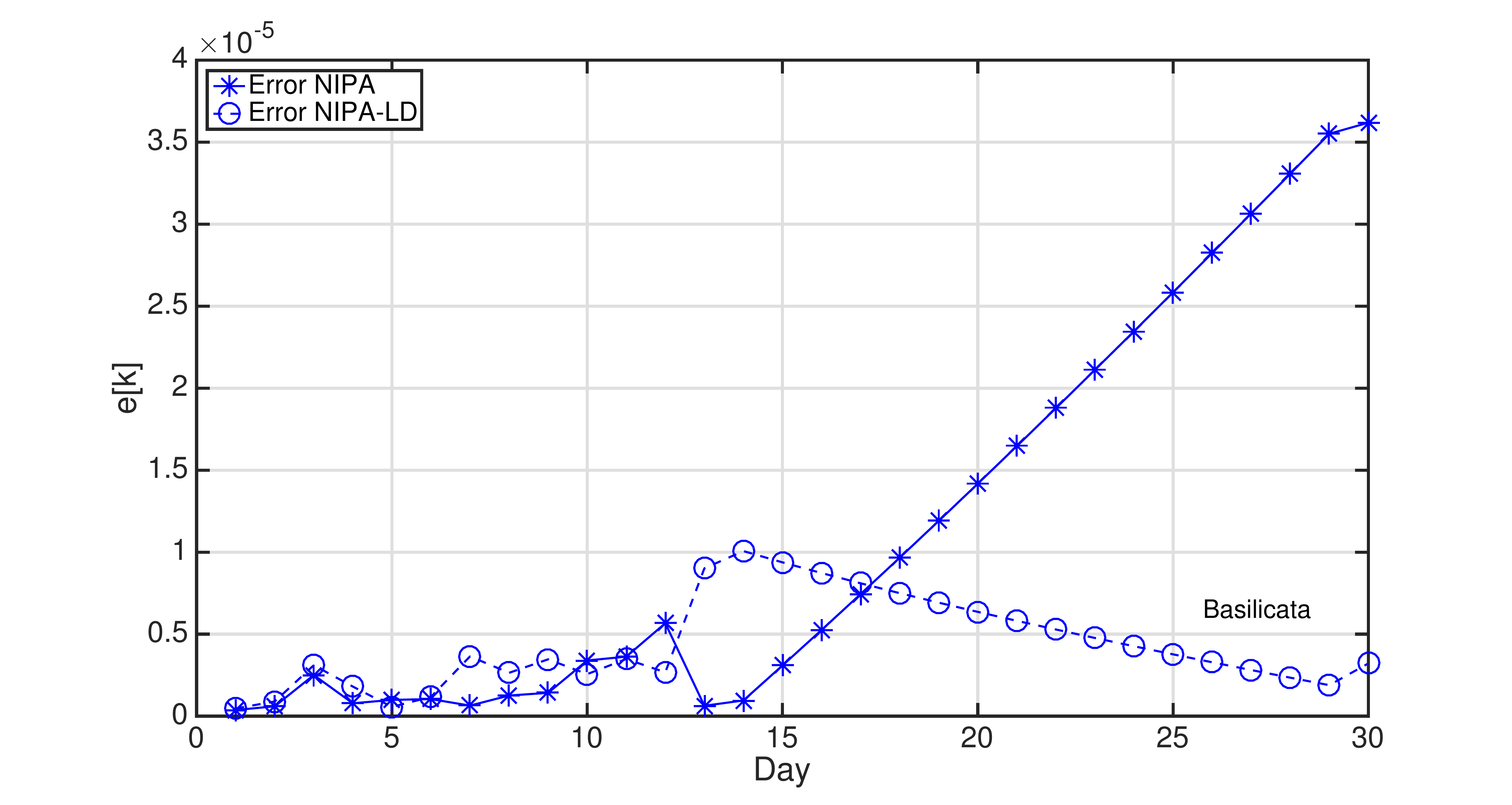}
	     \label{fig:Bas}}
\subfigure[]{
	   \includegraphics[width=0.45\linewidth]{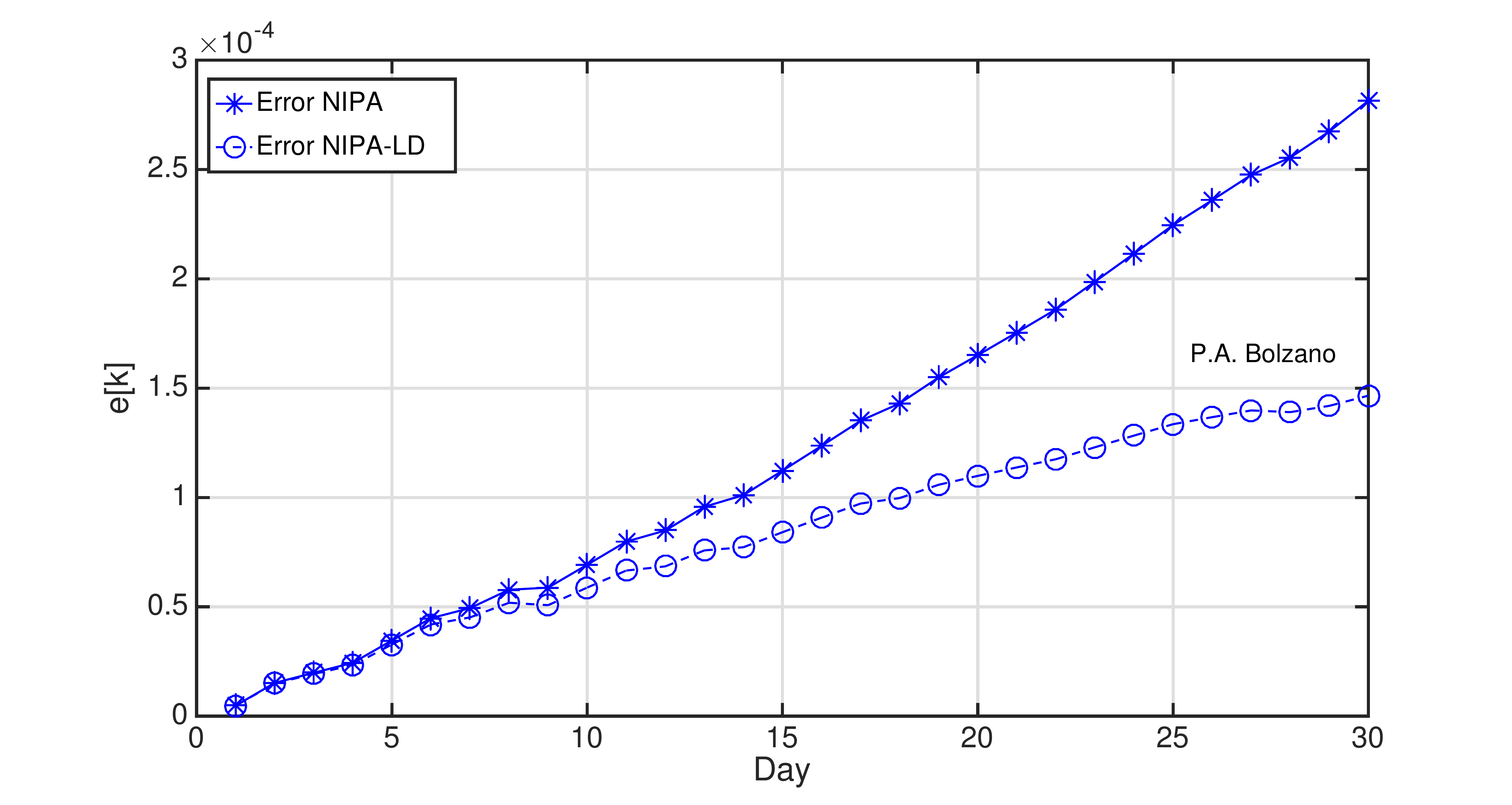}
	     \label{fig:Bol}}
\subfigure[]{
	   \includegraphics[width=0.45\linewidth]{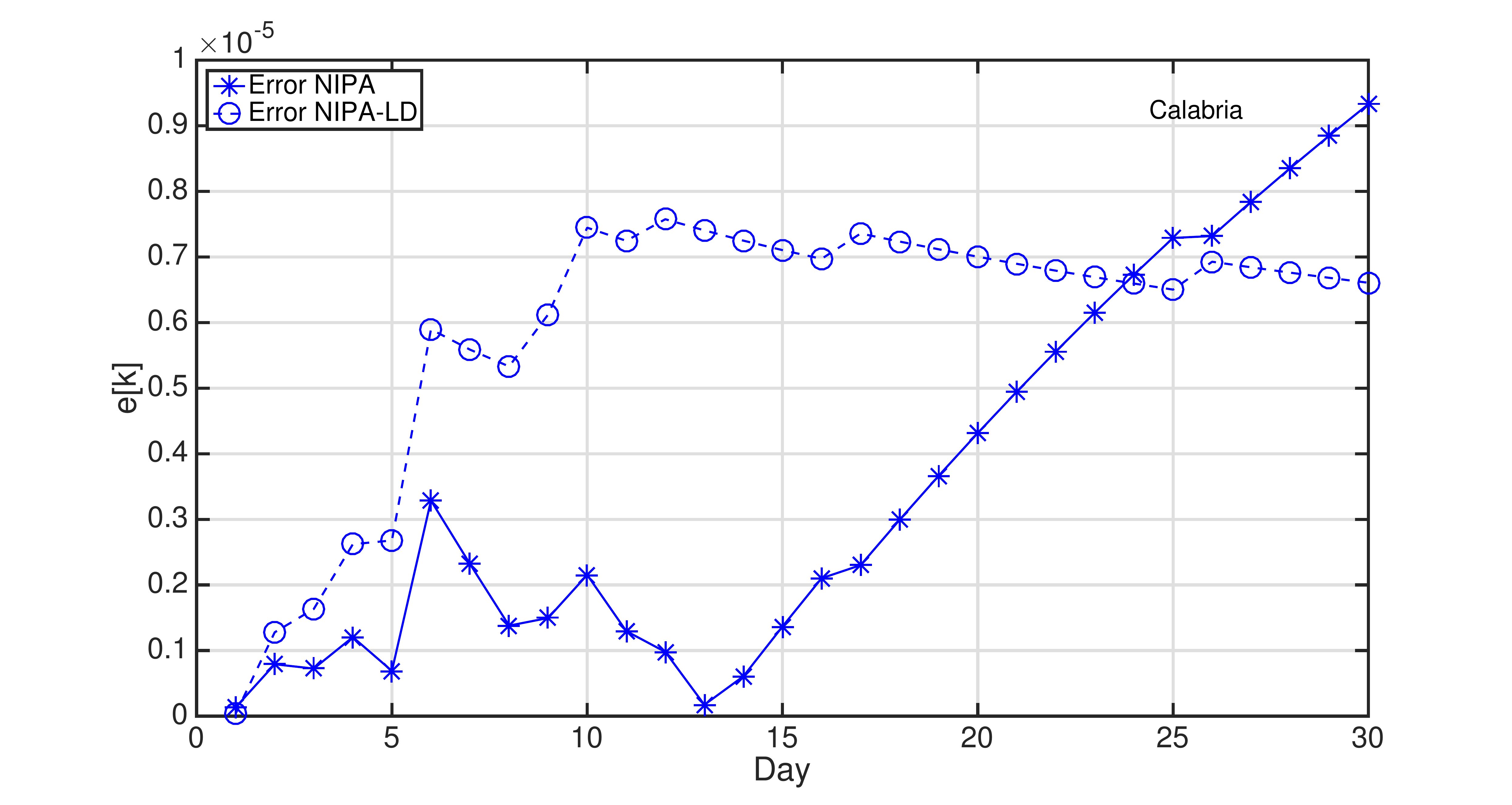}
	     \label{fig:Cal}}
\subfigure[]{
	   \includegraphics[width=0.45\linewidth]{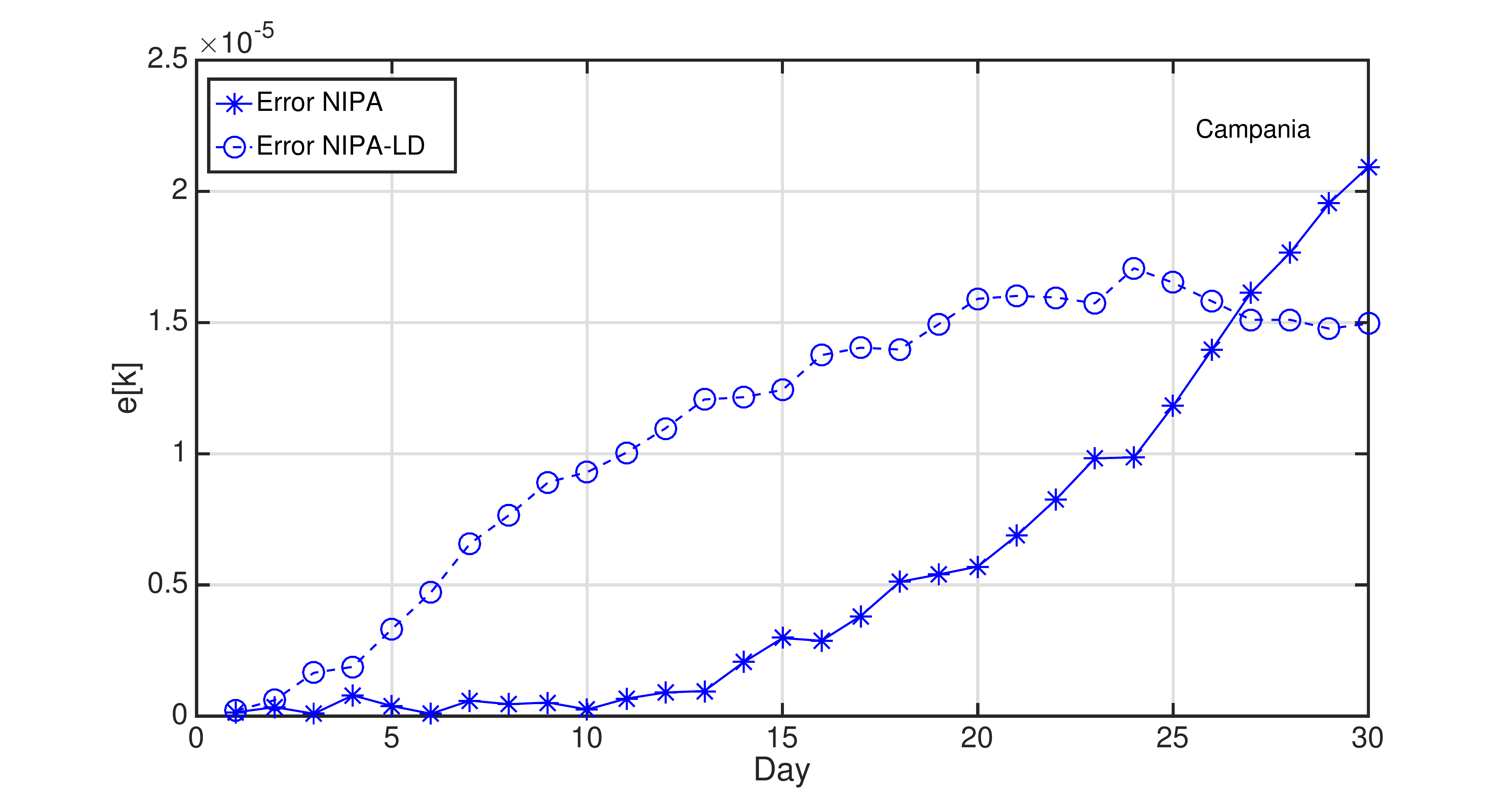}
	     \label{fig:Camp}}
\subfigure[]{
	   \includegraphics[width=0.45\linewidth]{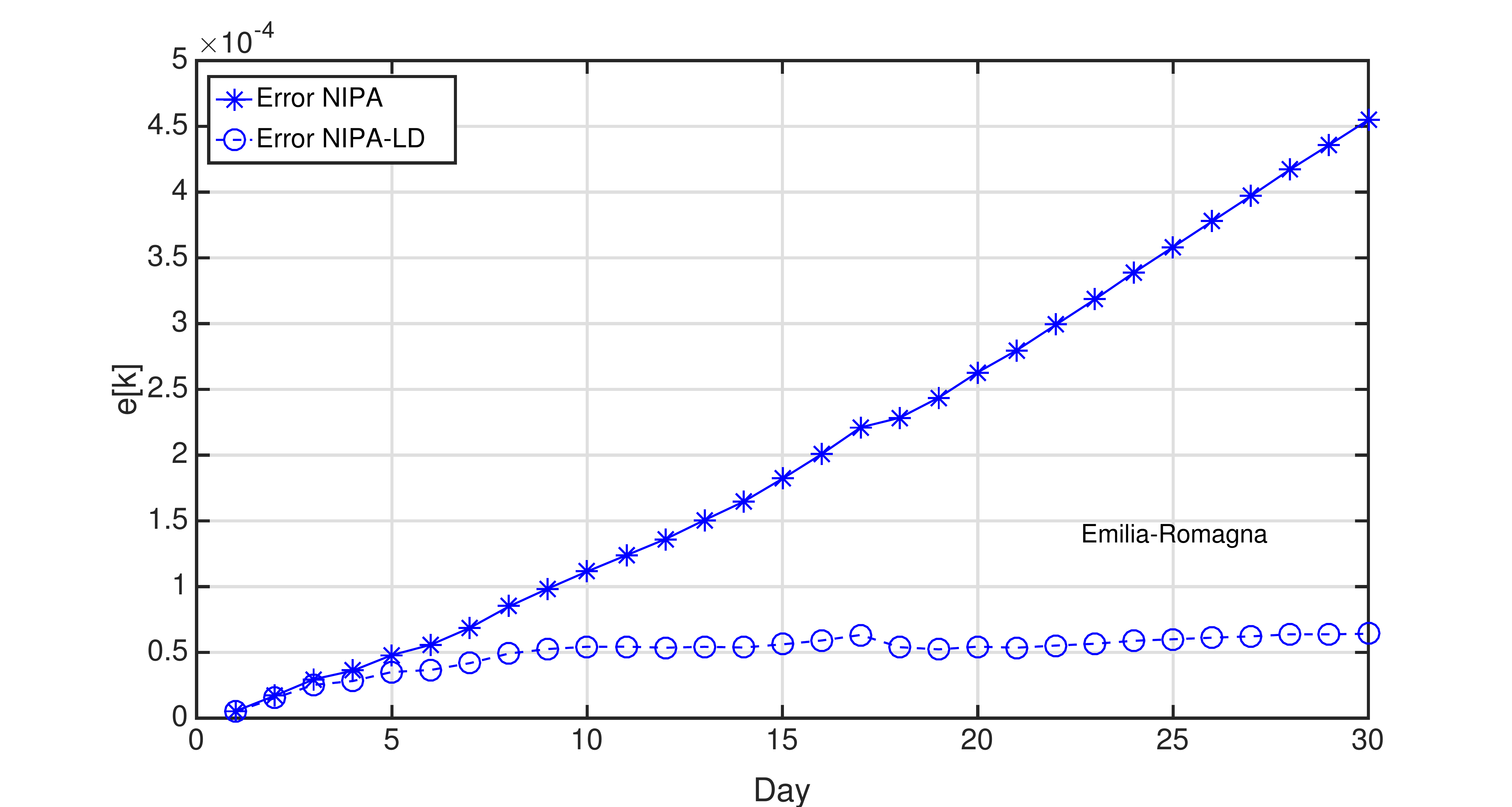}
	     \label{fig:Em}}
\subfigure[]{
	   \includegraphics[width=0.45\linewidth]{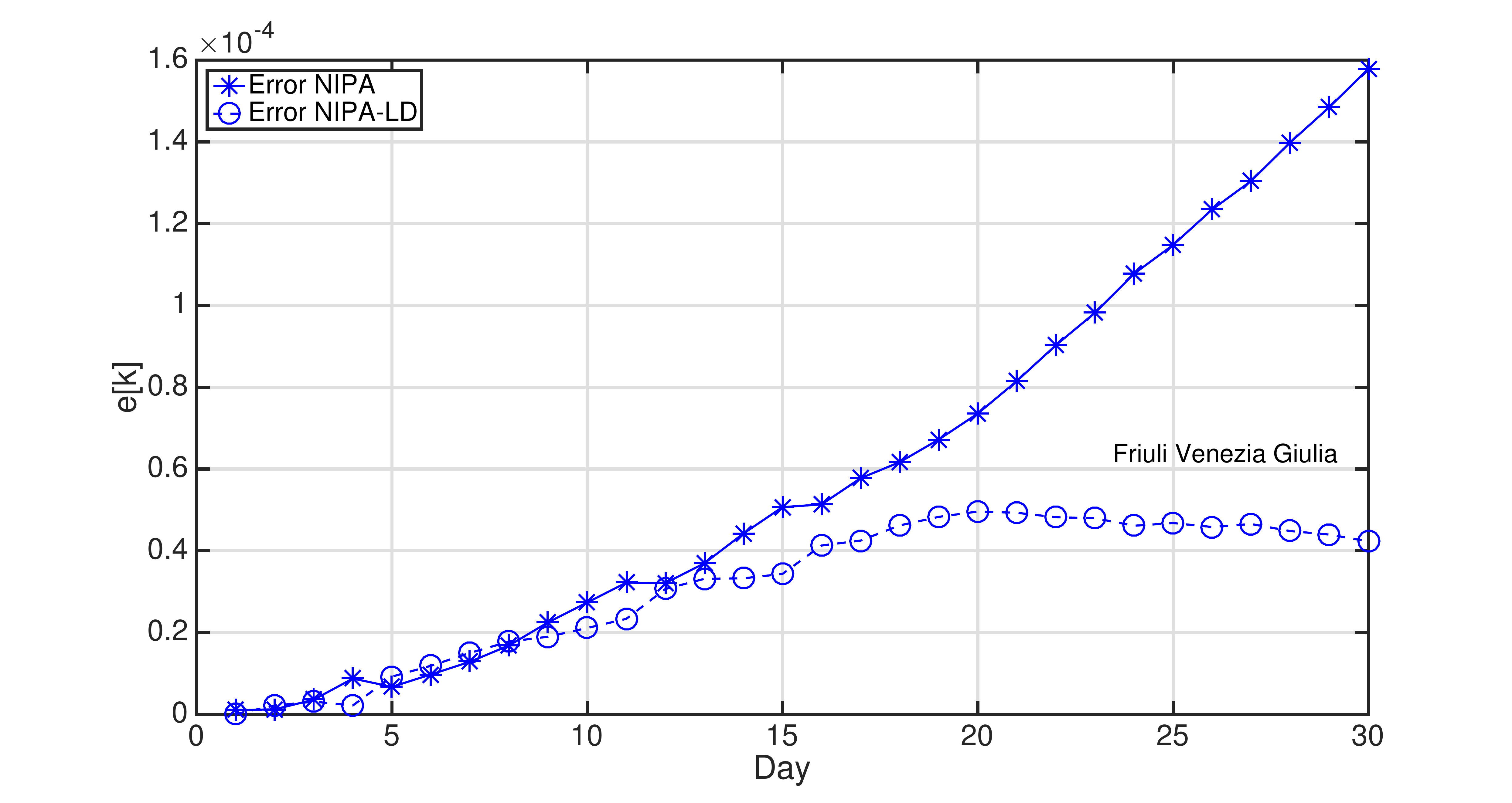}
	     \label{fig:Fri}}
\caption{Mean relative prediction error for the period from May 10 to June 9: first group of regions.}
\label{fig:reg_err_1_a}
\end{figure}

\begin{figure}	
\centering	     
\subfigure[]{
	   \includegraphics[width=0.45\linewidth]{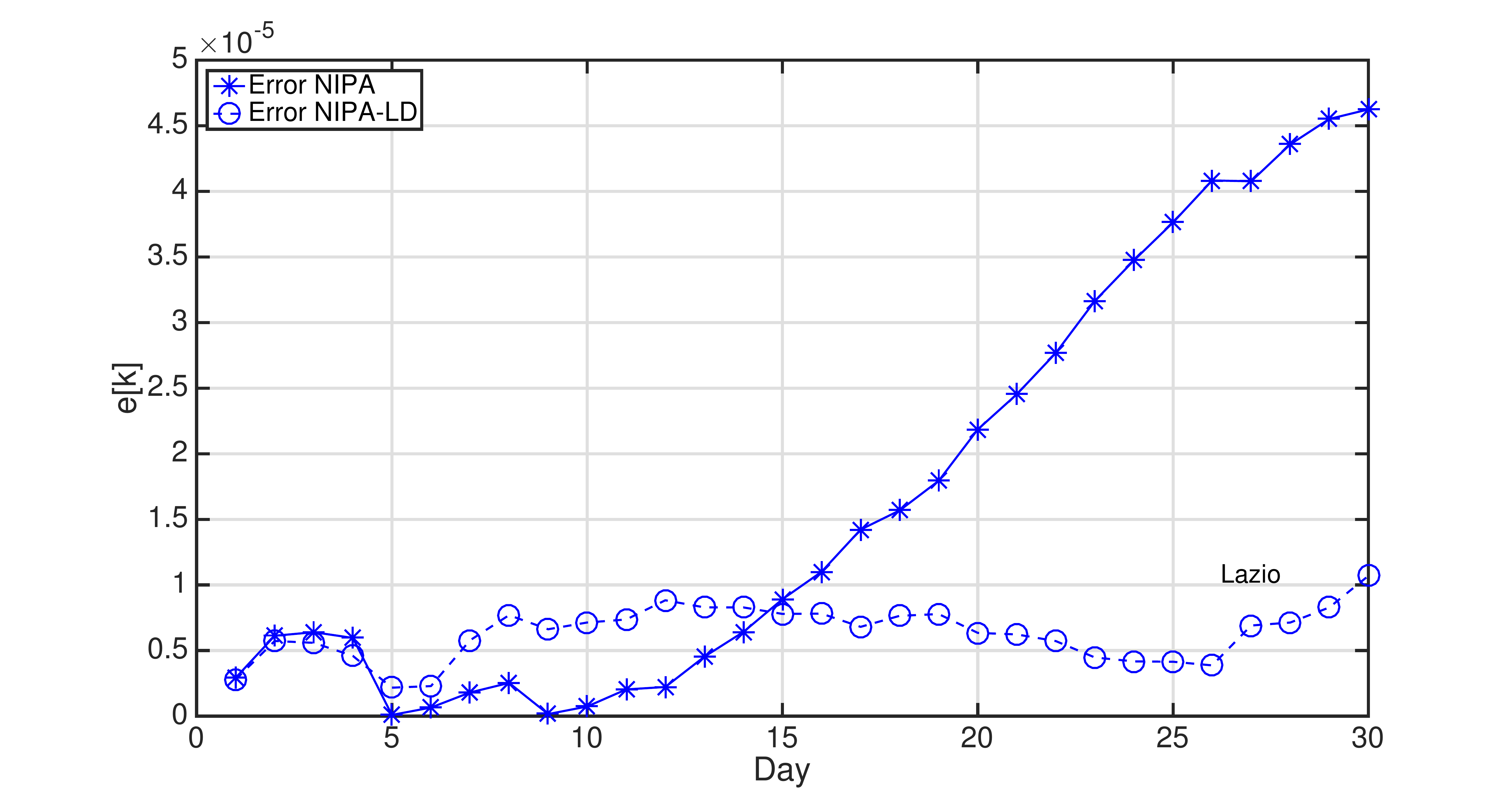}
	     \label{fig:Laz}}
\subfigure[]{
	   \includegraphics[width=0.45\linewidth]{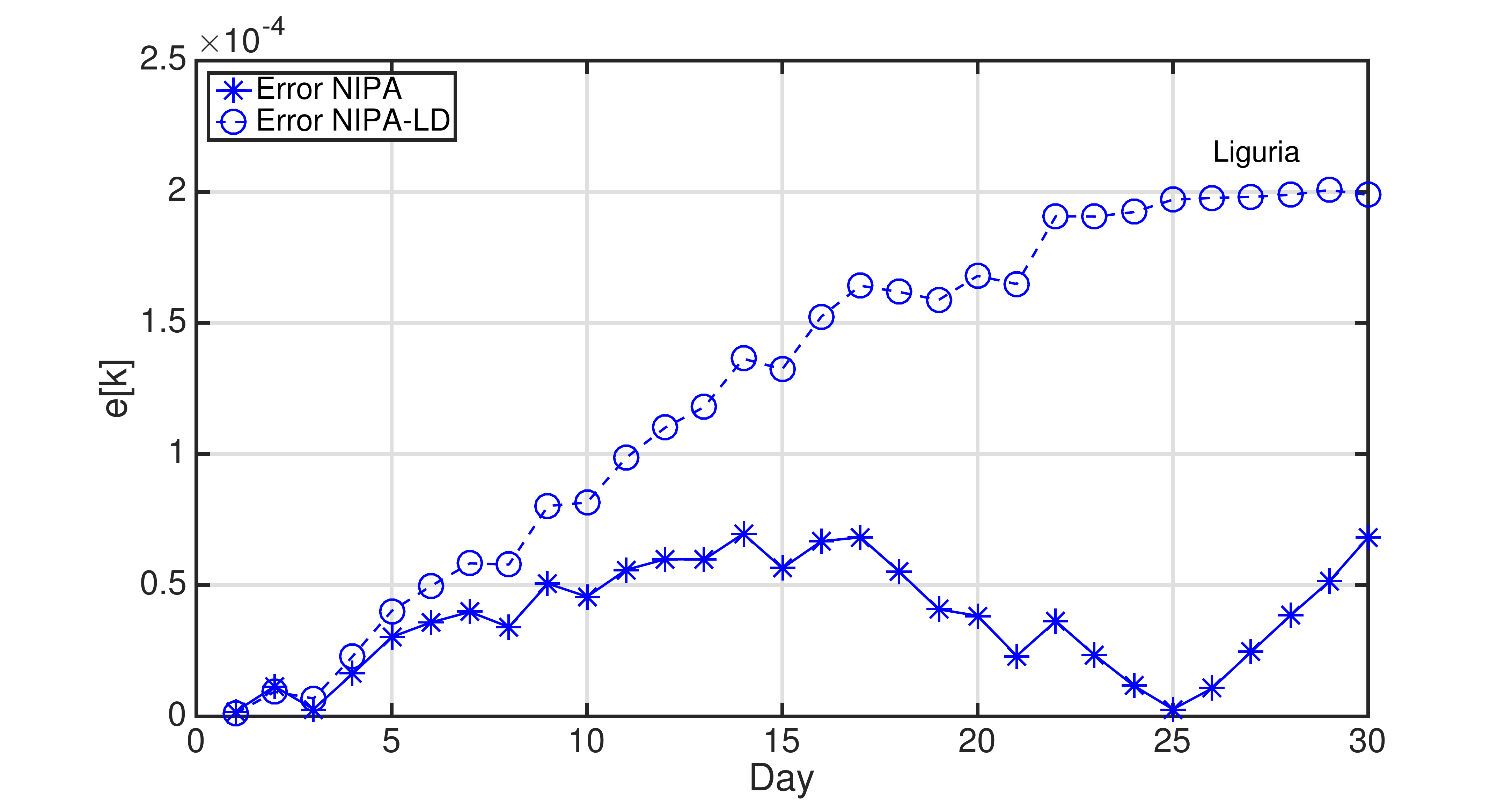}
	     \label{fig:Lig}}
\subfigure[]{
	   \includegraphics[width=0.45\linewidth]{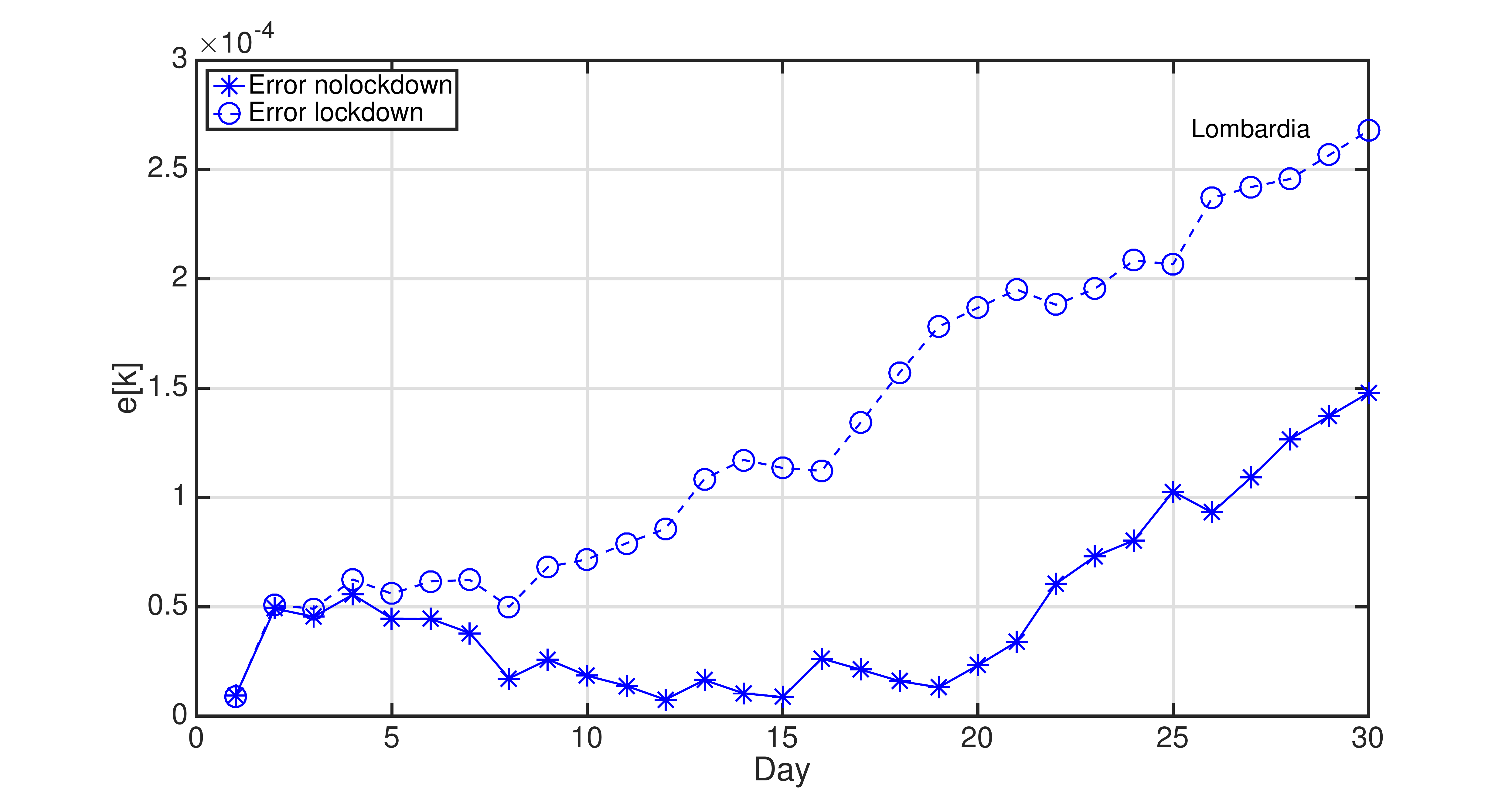}
	     \label{fig:Lom}}
\subfigure[]{
	   \includegraphics[width=0.45\linewidth]{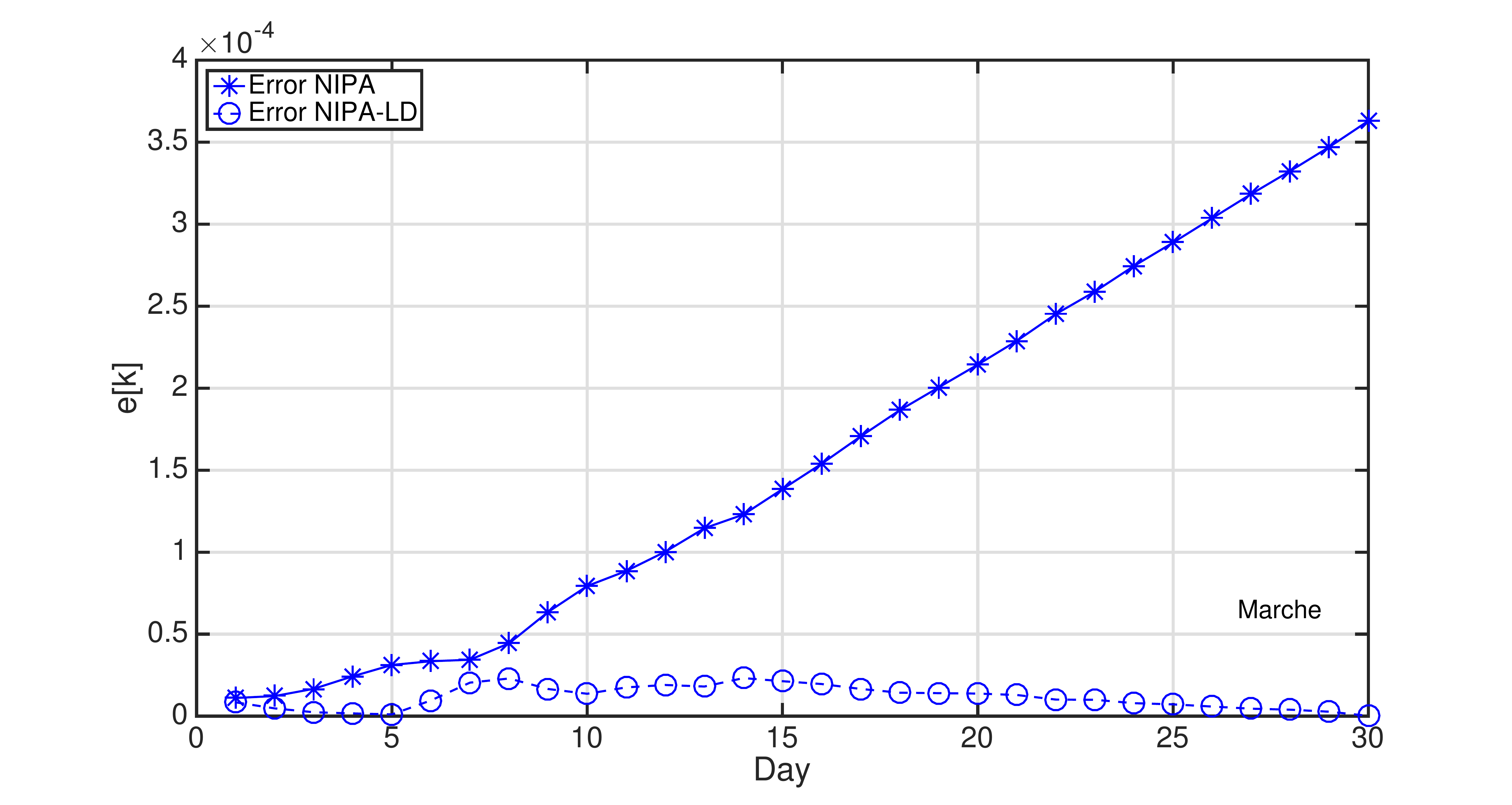}
	     \label{fig:Mar}}
\subfigure[]{
	   \includegraphics[width=0.45\linewidth]{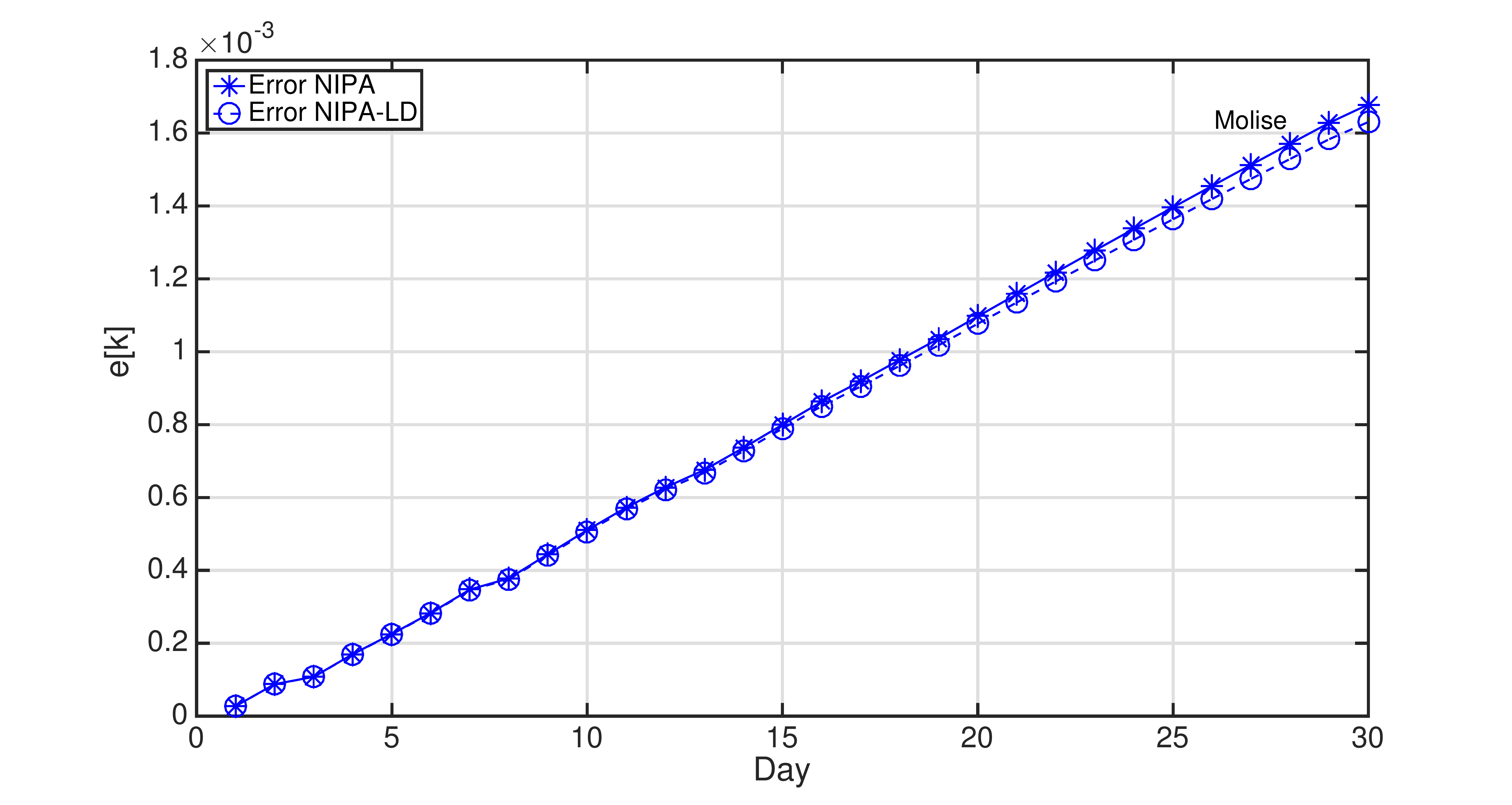}
	     \label{fig:Mol}}
\subfigure[]{
	   \includegraphics[width=0.45\linewidth]{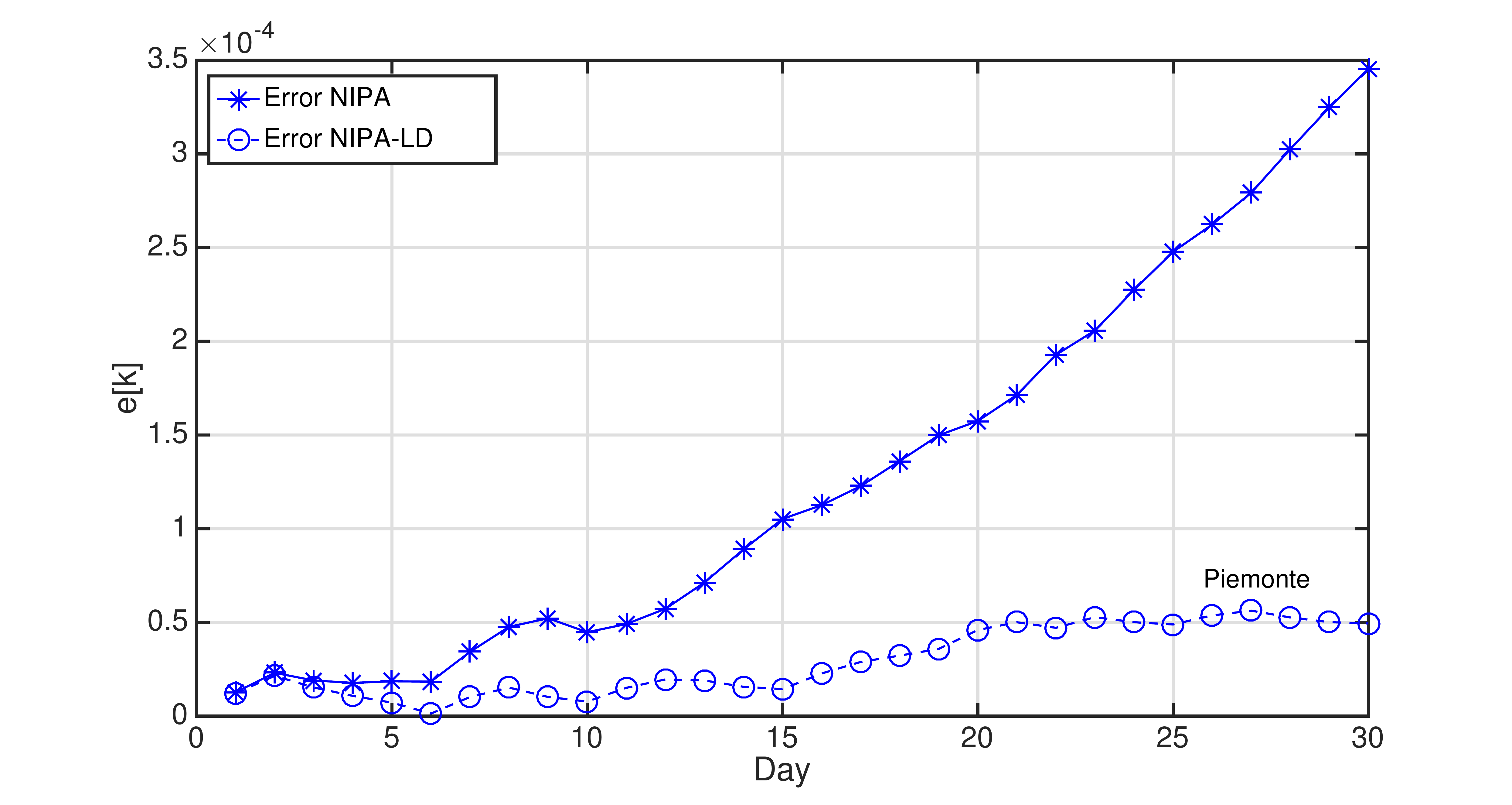}
	     \label{fig:Pie}}
\subfigure[]{
	   \includegraphics[width=0.45\linewidth]{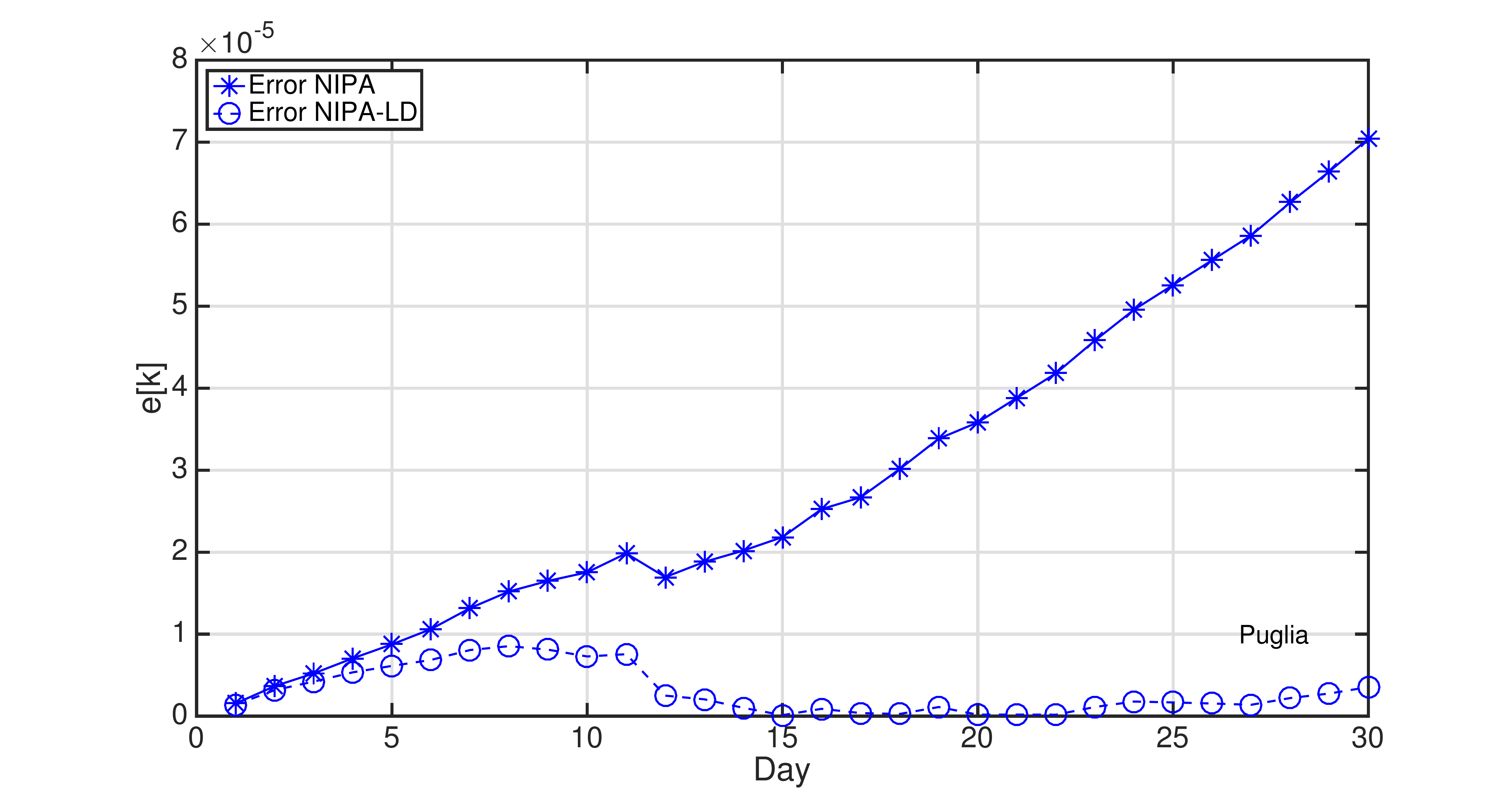}
	     \label{fig:Pu}}
\caption{Mean relative prediction error for the period from May 10 to June 9: second group of regions.}
\label{fig:reg_err_1_b}
\end{figure}

\begin{figure}	
\centering
\subfigure[]{
	   \includegraphics[width=0.45\linewidth]{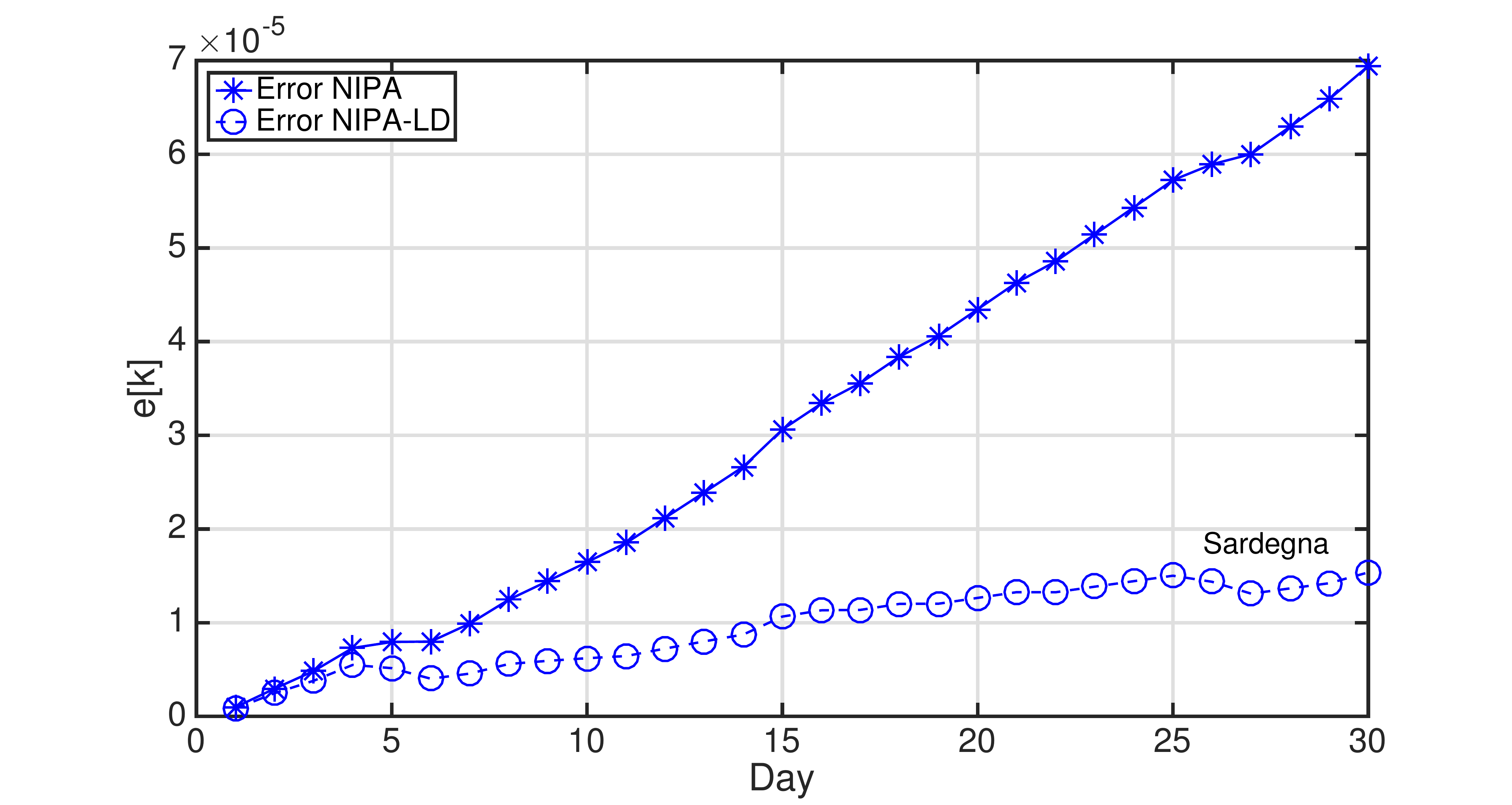}
	     \label{fig:Sar}}
\subfigure[]{
	   \includegraphics[width=0.45\linewidth]{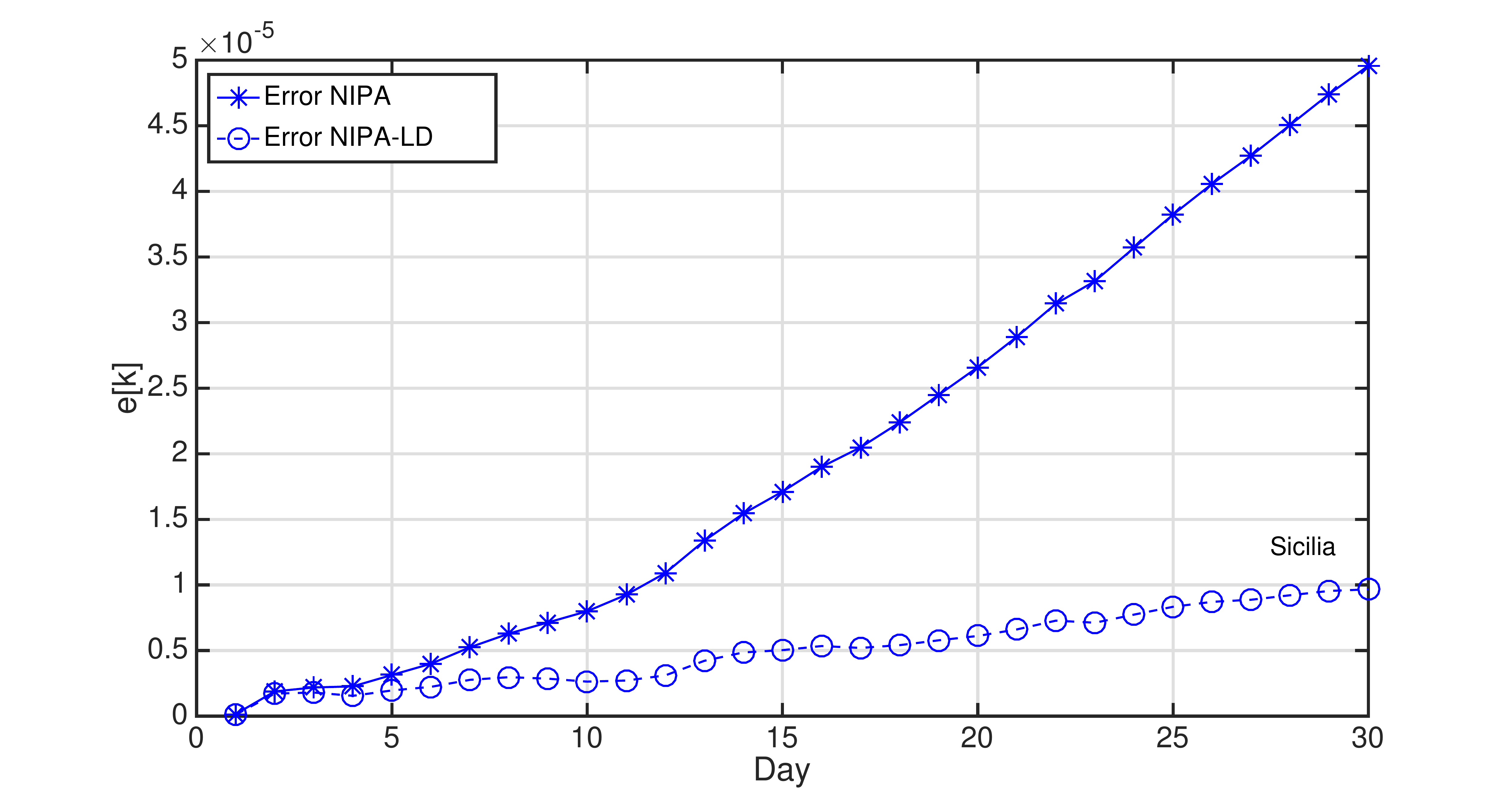}
	     \label{fig:Sic}}
\subfigure[]{
	   \includegraphics[width=0.45\linewidth]{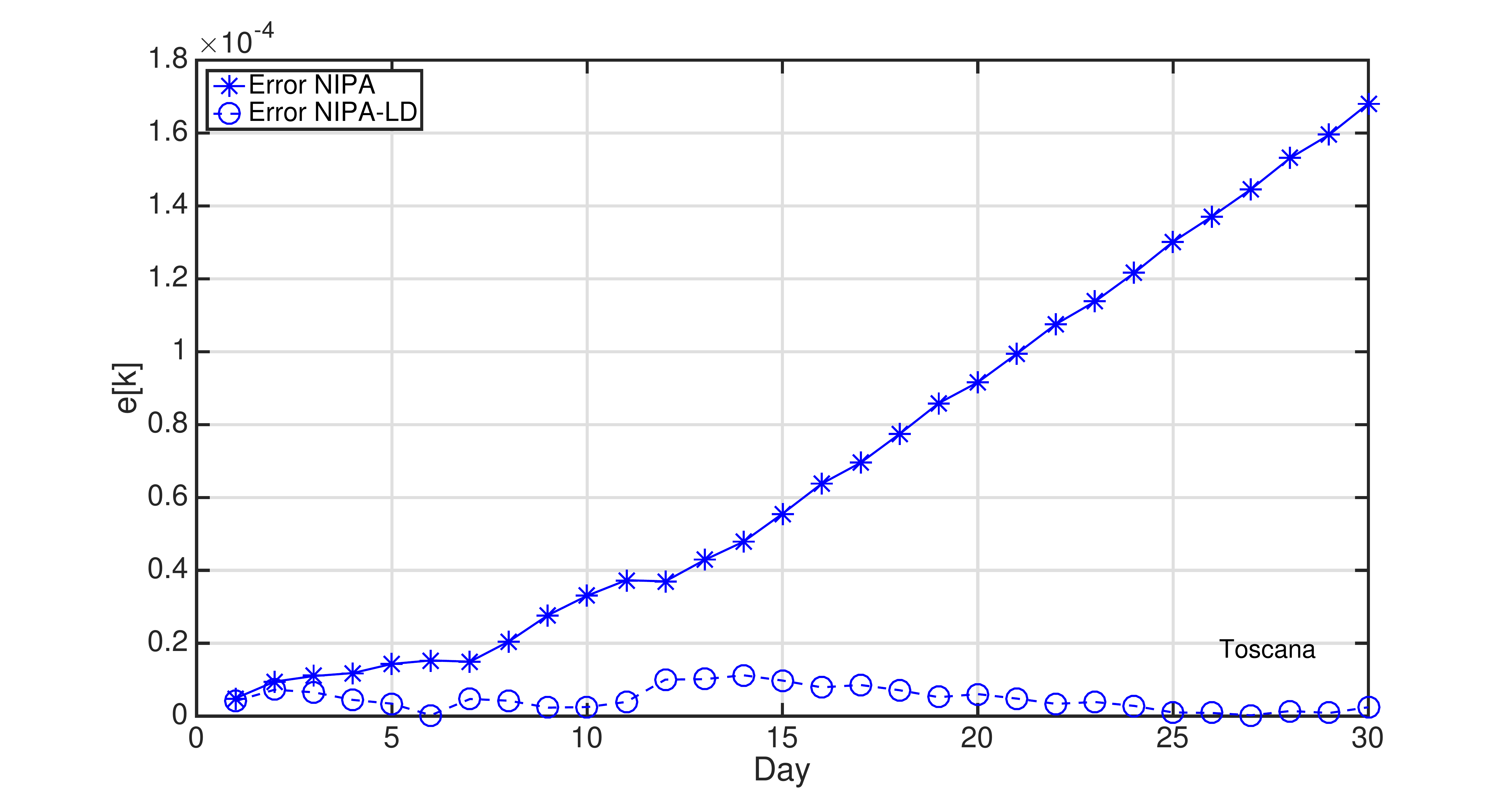}
	     \label{fig:Tos}}
\subfigure[]{
	   \includegraphics[width=0.45\linewidth]{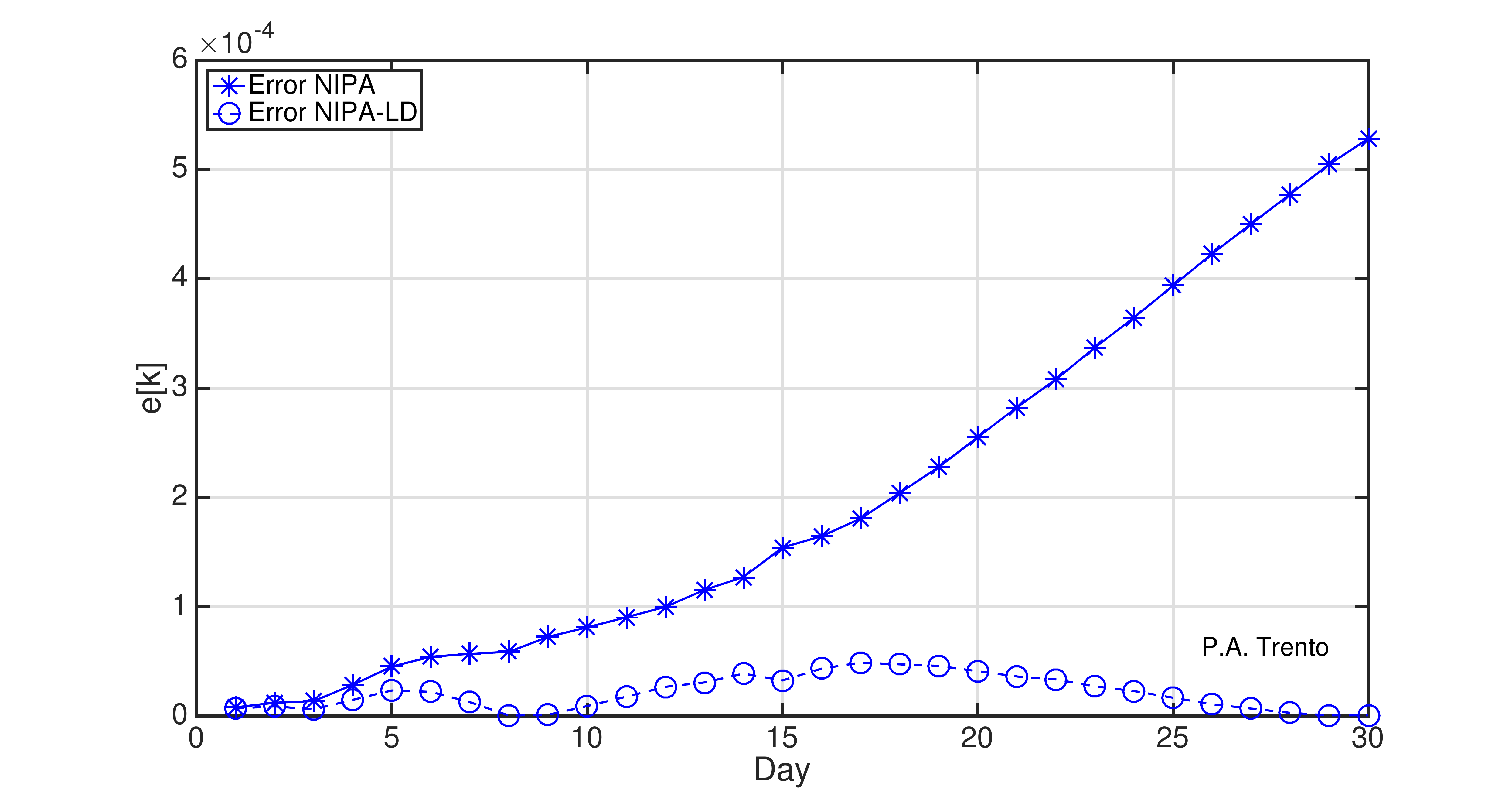}
	     \label{fig:Tren}}
\subfigure[]{
	   \includegraphics[width=0.45\linewidth]{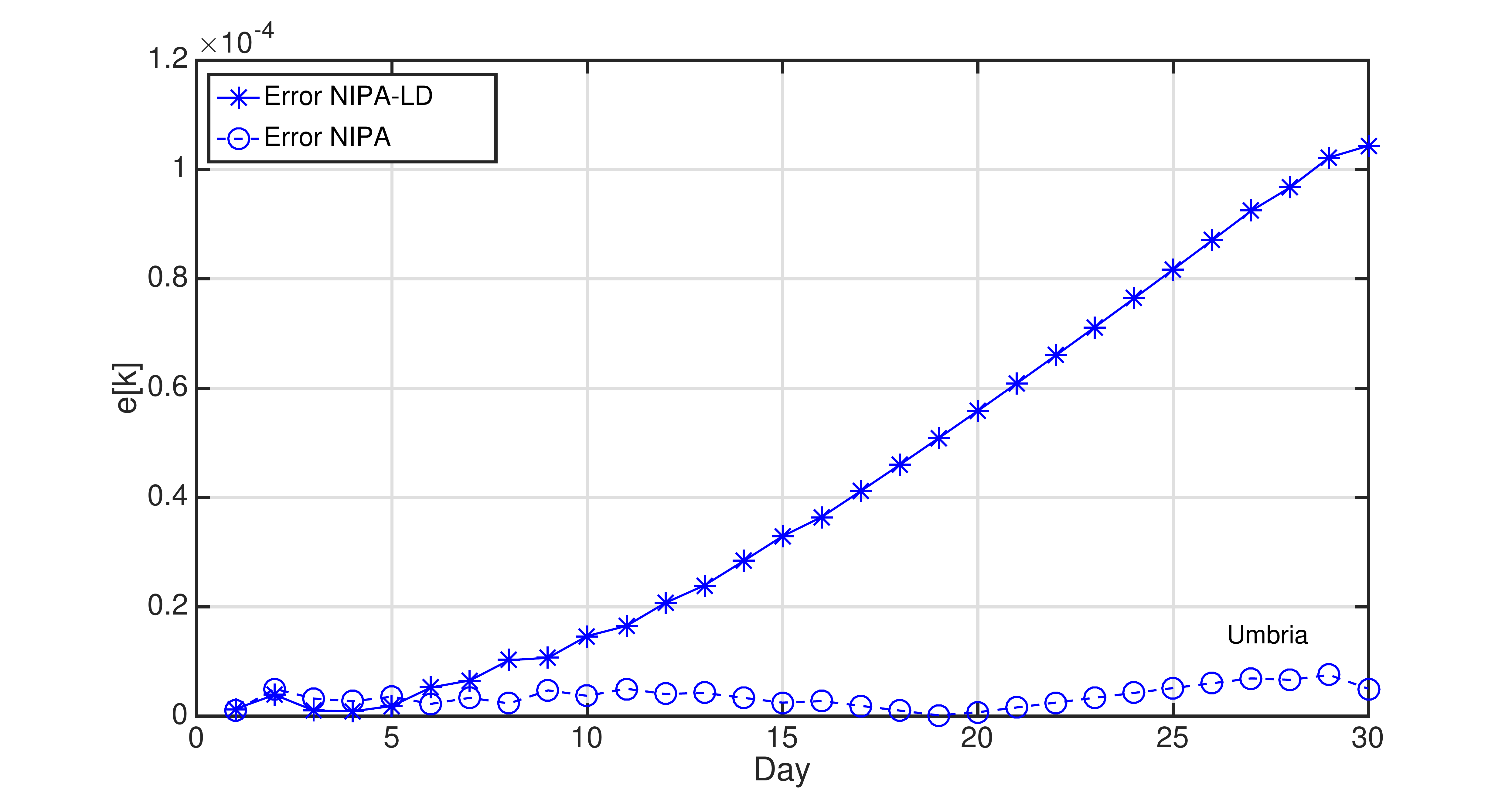}
	     \label{fig:Um}}
\subfigure[]{
	   \includegraphics[width=0.45\linewidth]{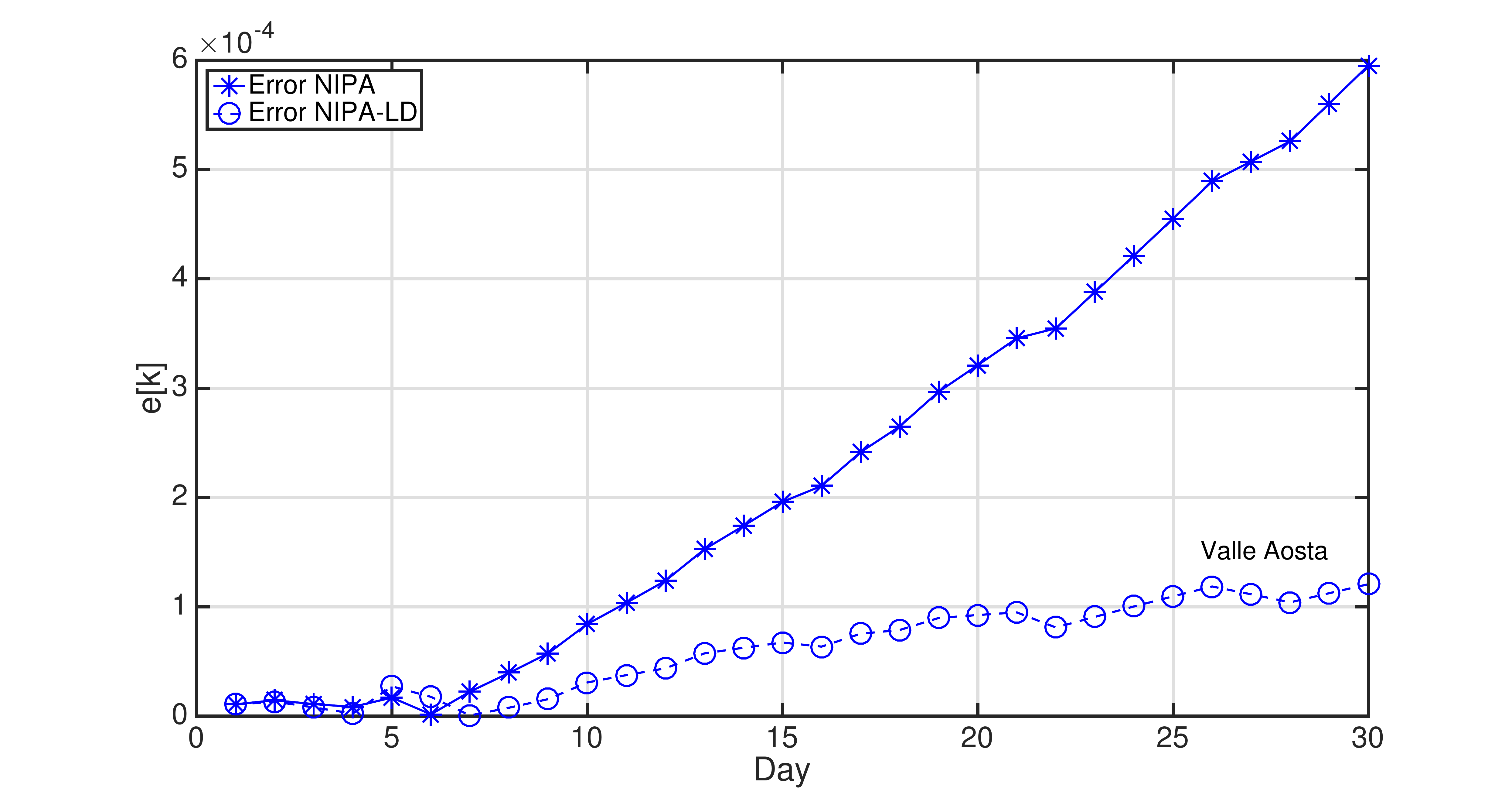}
	     \label{fig:Aos}}
\subfigure[]{
	   \includegraphics[width=0.45\linewidth]{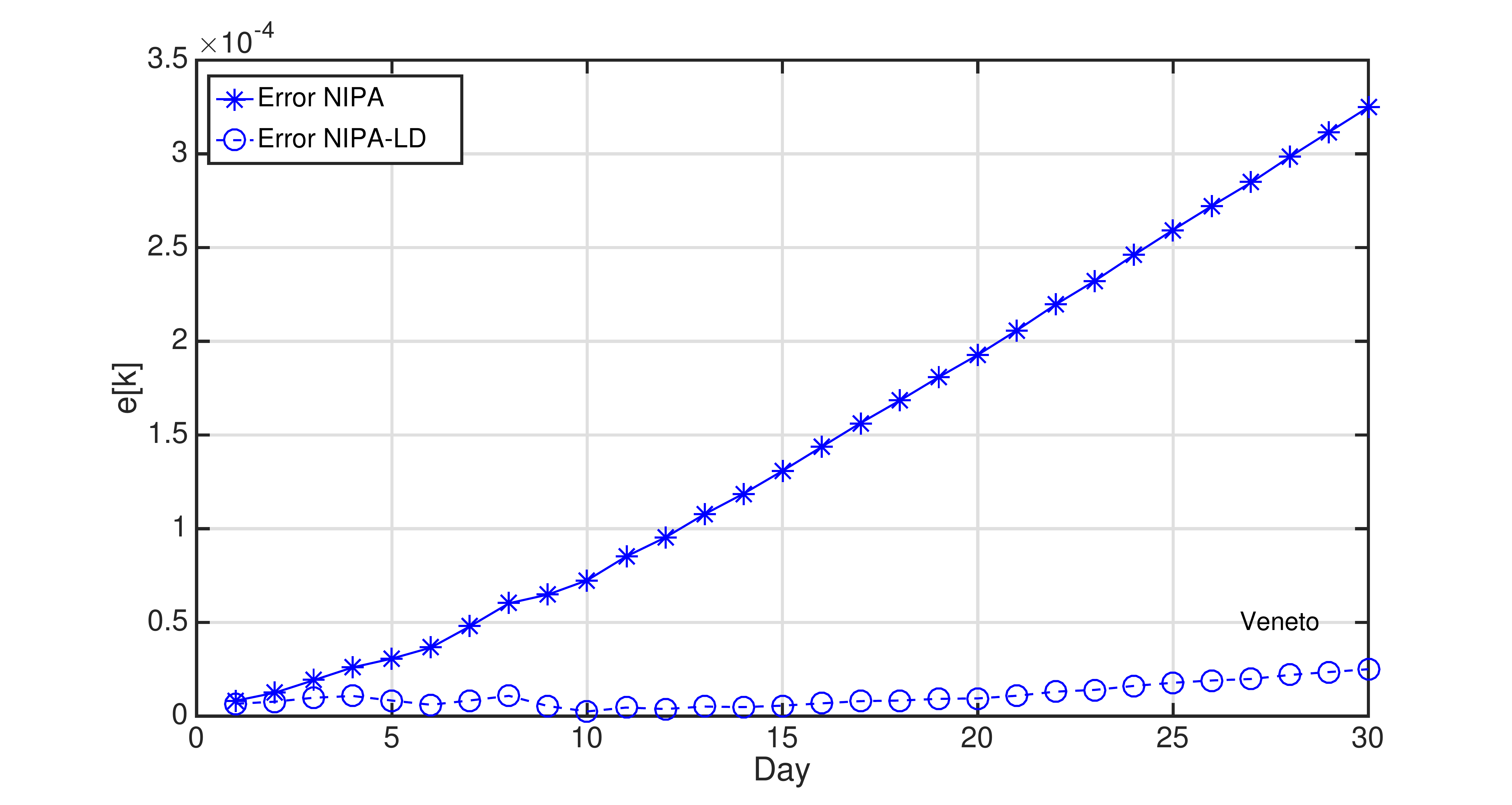}
	     \label{fig:Ven}}	     	     	     
\caption{Mean relative prediction error for the period from May 10 to June 9: third group of regions.}
\label{fig:reg_err_2}
\end{figure}

Figures \ref{fig:reg_err_1_a}, \ref{fig:reg_err_1_b}  and \ref{fig:reg_err_2} report the mean relative prediction error $e[k]$ for the 21 regions divided into 3 groups, over an observation period of 30 days from May 10 to June 9\asf{. For} most of the regions (\textit{P.A. Bolzano}, \textit{Emilia-Romagna}, \textit{Friuli Venezia Giulia}, \textit{Marche}, \textit{Piemonte}, \textit{Puglia}, \textit{Sardegna}, \textit{Sicilia}, \textit{Toscana}, \textit{P.A. Trento}, \textit{Umbria}, \textit{Valle d'Aosta}, \textit{Veneto}) 
 NIPA-LD results in a substantially lower prediction error. In particular, after few days the re-openings of May 18 (corresponding to the third day in the plots), for which the population gradually started again going to bars, shops, hair dressers and other commercial activities and exploiting other kind of allowed services, the prediction error is much lower with the lockdown applied to NIPA. In other regions, like \textit{Abruzzo}, \textit{Basilicata}, \textit{Calabria}, \textit{Campania}, and \textit{Lazio}, NIPA performs better than NIPA-LD for many days after May 16. This behavior could be
 due to the fact that on May 18 the mobility among the Italian region was allowed, thus there has been a high flow of people moving towards the southern regions. Thus, in spite of the restrictions made by the regional governor, often much more strict than the national ones, like, for instance in Campania,  the lockdown measures where not effective. For \textit{Liguria} and \textit{Lombardia}, characterized by much more COVID-19 cases compared to the other regions, NIPA results in a lower error. Also for these two regions it seems that lockdown measures did not work. Finally, the \textit{Molise} case is the only one having no substantial difference between the prediction error with lockdown and without lockdown. This region had the lowest number of COVID-19 cases. Moreover, there has been an erratic change in the number of infections in \asf{\textit{Molise}}, due to a single group of people, who did not follow the quarantine measures imposed by the Italian Government. 
 
\begin{figure}	
\centering
\includegraphics[width=0.7\linewidth]{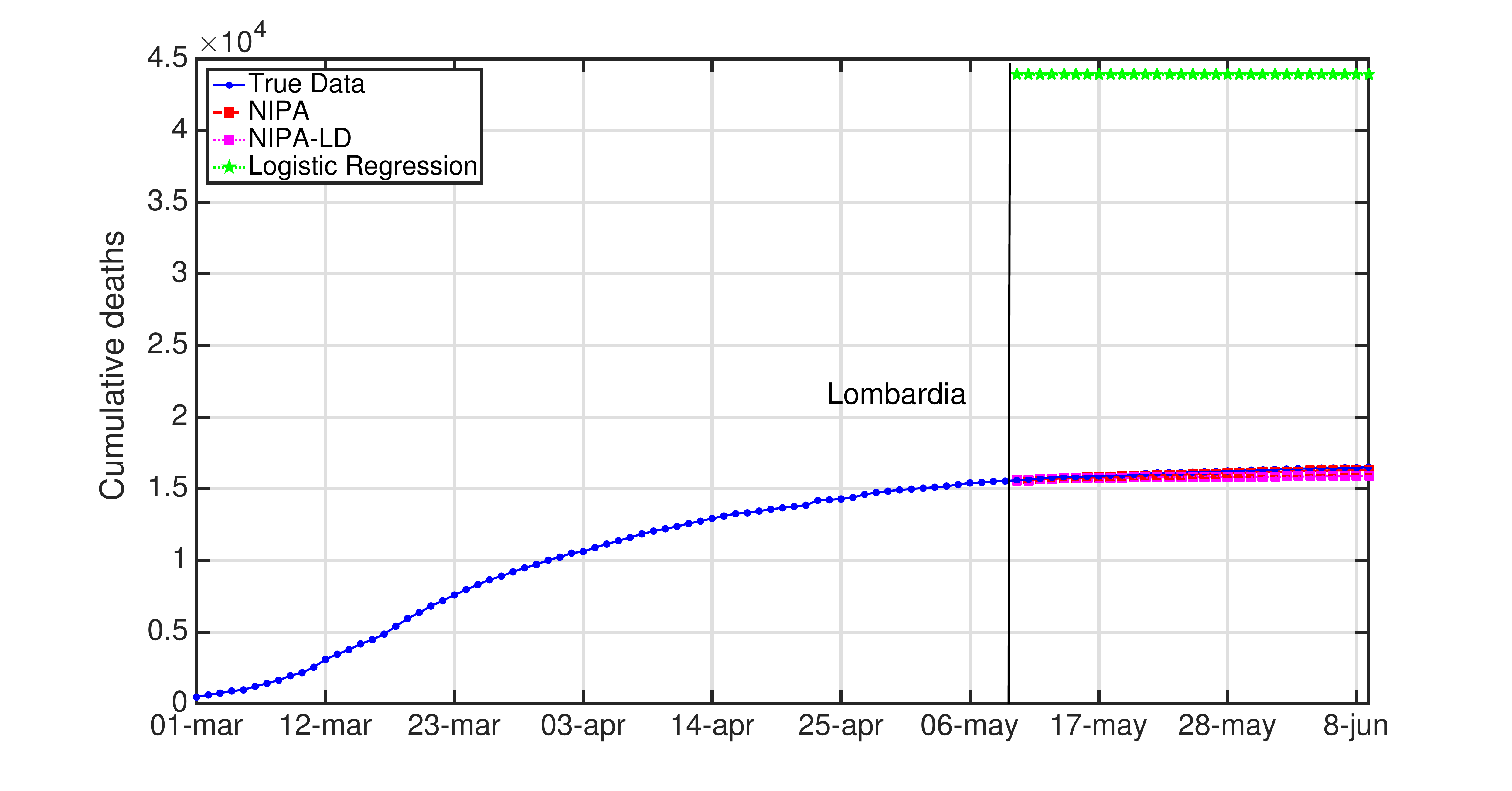}
\caption{Cumulative deaths for \textit{Lombardia} with $n_{neglect}=30$.}
\label{fig:deathLom}
\end{figure}

\as{\section*{Death Prediction}
The network-based SIR model described in the paper does not consider the death cases. To predict the number of deaths a new compartment should be added. However, by substituting the cumulated cases of infected with those of
dead people, the model allows to predict the deaths. Thereby, we assume that the number of \asf{deaths} is proportional to the number of infections. Thus, we executed NIPA and NIPA-LD on these cumulated death cases to predict the deaths instead of the infections. Even if 
the death numbers are subject to greater variations and there are significantly fewer deaths than infections, the methods give good results. 
Table \ref{tab:deaths}  reports for each region the average MAPE error for NIPA and NIPA-LD in predicting COVID-19 deaths. For this experiment we set the number of neglected days to 30 by using the same transmission modifier values of the previous experiments.  The table shows that the error values are very low and  that NIPA-LD outperforms NIPA in 14 out of the 21 regions.  It is worth pointing out that when NIPA performs better, the differences between error values are very low, except for \asf{\textit{Lombardia}}. As known, this region had more than 16 thousands deaths in the considered period.  NIPA-LD in such a case underestimates the number of deaths. Figure \ref{fig:deathLom} shows the predicted cumulative deaths of these two methods and those predicted by using logistic regression. Note that the baseline function is not able to obtain a good prediction, in fact it overestimates too much the number of deaths.}

\begin{table}
\centering
%\scriptsize
%\footnotesize
\centering
%\tiny
\centering
\caption{\as{Average MAPE prediction error of COVID-19 deaths for NIPA and NIPA-LD, when the number of neglected days is 30.}}
\label{tab:deaths}
\begin{tabular}{|c|c|c|} \hline
REGION & Error NIPA & Error NIPA-LD\\\hline
$Abruzzo$ & {\bf 0.1065*10$^{-4}$}  &  0.1293*10$^{-4}$ \\\hline
$Basilicata$ & 0.0198*10$^{-4}$  &  {\bf 0.0114*10$^{-4}$} \\\hline
$P.A. Bolzano$ & 0.0714*10$^{-4}$ &   {\bf 0.0204*10$^{-4}$} \\\hline
$Calabria$ & 0.0051*10$^{-4}$ &   {\bf 0.0031*10$^{-4}$} \\\hline
$Campania$ &  {\bf 0.0133*10$^{-4}$}   & 0.0233*10$^{-4}$\\\hline
$Emilia$ & {\bf 0.0244*10$^{-4}$}  &  0.2203*10$^{-4}$\\\hline
$Friuli$ & {\bf 0.0846*10$^{-4}$}  &  0.1081*10$^{-4}$\\\hline
$Lazio$ & {\bf 0.0314*10$^{-4}$}   & 0.0416*10$^{-4}$\\\hline
$Liguria$ & 0.1840*10$^{-4}$  &  {\bf 0.1562*10$^{-4}$}\\\hline
$Lombardia$ & {\bf 0.133*10$^{-4}$}  &  0.3821*10$^{-4}$\\\hline
$Marche$ & 0.1006*10$^{-4}$  &  {\bf 0.0621*10$^{-4}$}\\\hline
$Molise$ & 0.3505*10$^{-4}$ &   {\bf 0.0931*10$^{-4}$}\\\hline
$Piemonte$ & 0.3525*10$^{-4}$   & {\bf 0.2932*10$^{-4}$}\\\hline
$Puglia$ & 0.3811*10$^{-4}$ &   {\bf 0.3170*10$^{-4}$}\\\hline
$Sardegna$ & 0.0102*10$^{-4}$  &  {\bf 0.0036*10$^{-4}$}\\\hline
$Sicilia$ & 0.0090*10$^{-4}$  &  {\bf 0.0017*10$^{-4}$}\\\hline
$Toscana$ & 0.0488*10$^{-4}$  &  {\bf 0.0675*10$^{-4}$}\\\hline
$P.A. Trento$ & 0.1106*10$^{-4}$  &  {\bf 0.0142*10$^{-4}$}\\\hline
$Umbria$ & {\bf 0.0038*10$^{-4}$}  &  0.0089*10$^{-4}$\\\hline
$Valle Aosta$ & 0.0493*10$^{-4}$  &  {\bf 0.0335*10$^{-4}$}\\\hline
$Veneto$ & 0.0423*10$^{-4}$  &  {\bf 0.0055*10$^{-4}$}\\
\hline
\end{tabular}
\end{table}

\section*{Discussion}

The results reported in the previous section  show that NIPA-LD is able to better predict the evolution of COVID-19 in Italy when compared to the original NIPA method, that does not consider the lockdown measures, and to the baseline prediction method. The main contribution of NIPA-LD is the capability of sensibly improving the long-term prediction of NIPA by implementing  the different lockdown measures adopted in the various phases of the spreading of the COVID-19 in Italy  into the network-based prediction model. In fact, NIPA-LD obtains lower prediction errors than NIPA when the number of training days diminishes. The introduction of the concept of transmission modifiers in NIPA thus allows to have epidemic transmission rates which well reflect the changes in the containment measured imposed by authorities. 

However, the adoption of the same values of transmission modifier for all the Italian regions has some drawbacks. In Tables \ref{tab:fv1p} and \ref{tab:fv2}, we report the daily error fraction value between  NIPA-LD and NIPA for 30 neglected days. In the last column of Table \ref{tab:fv2}, the average value of this error is also shown. When NIPA-LD outperforms NIPA, the daily error fraction is lower than 1. For most of the regions, NIPA-LD shows its superiority. \textit{Veneto}, for example, is characterized by very low values with an average daily error of 0.15. Exceptions are \textit{Abruzzo}, \textit{Basilicata}, \textit{Calabria}, \textit{Campania}, \textit{Lazio}, \textit{Liguria}, and \textit{Lombardia}, where NIPA performs better than NIPA-LD. Thus, though on average, NIPA-LD improves the prediction, this improvement is not for all the regions.   Future works will investigate specialized transmission modifiers for the different regions. 
Moreover, whereas the transmission modifier $\pi[k]$ may change over time, the infection rates $\beta_{ij}$ are assumed constant. Hence, in NIPA-LD (and classic NIPA) another limitation is that the probabilities of infection are assumed to be constant, or potentially scaled/multiplied by $\pi[k]$. Similarly, our model assumes constant curing rates $\delta$. However, (hopefully soon available) vaccinations may be deployed in a time-varying manner.

%\as{Another observation is that although NIPA and NIPA-LD can obtain good short-term predictions, accurate and real long-term predictions are generally not possible, regardless of the prediction algorithm. A thorough discussion in \cite{prasse2020fundamental}, shows how the prediction of epidemics is inherently difficult and not possible beyond some time horizons. The reason is that the logistic function usually exploited to describe the cumulative number of infections within the model is ill-conditioned.}

\begin{figure}	
\centering
\includegraphics[width=0.7\linewidth]{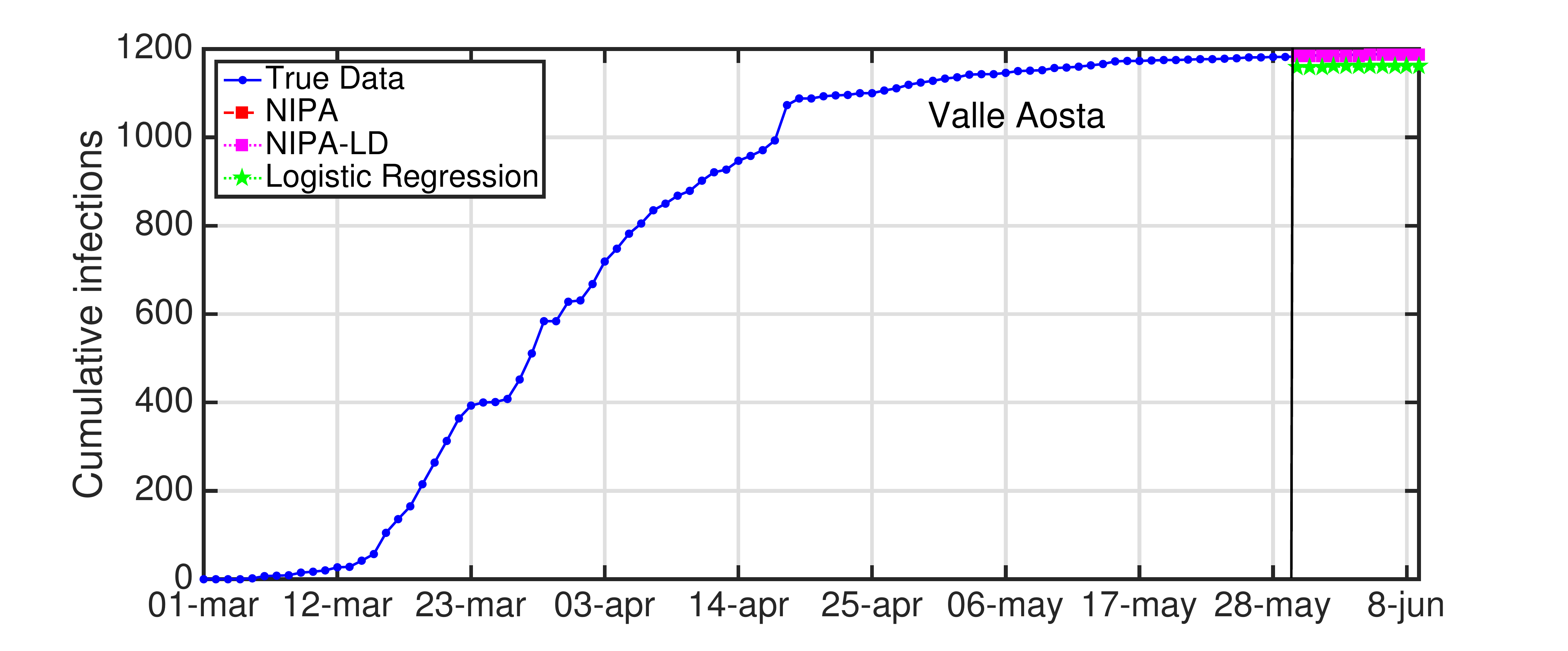}
\caption{Cumulative infections for \textit{Valle d'Aosta} with $n_{neglect}=10$.}
\label{fig:ValleA_1}
\end{figure}

\begin{figure}	
\centering
\includegraphics[width=0.7\linewidth]{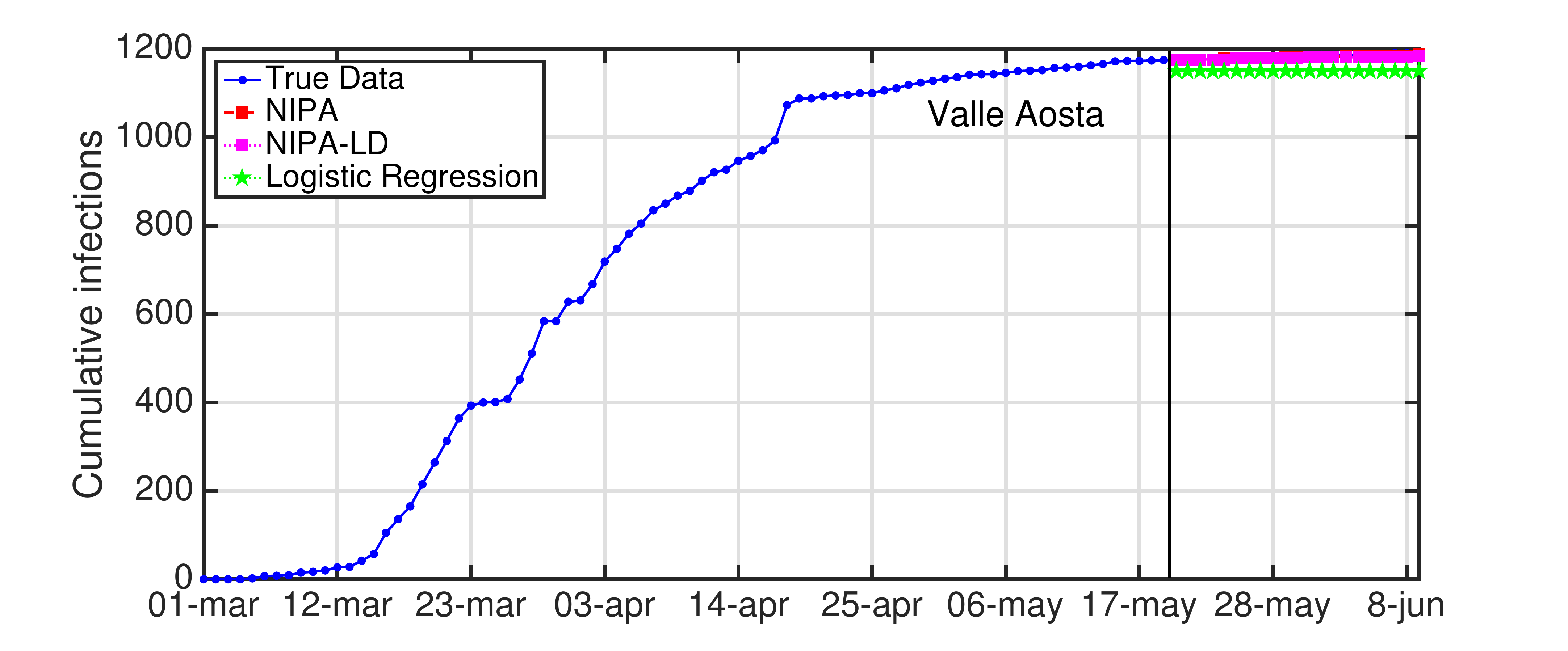}
\caption{Cumulative infections for \textit{Valle d'Aosta} with $n_{neglect}=20$.}
\label{fig:ValleA_2}
\end{figure}

\begin{figure}	
\centering
\includegraphics[width=0.7\linewidth]{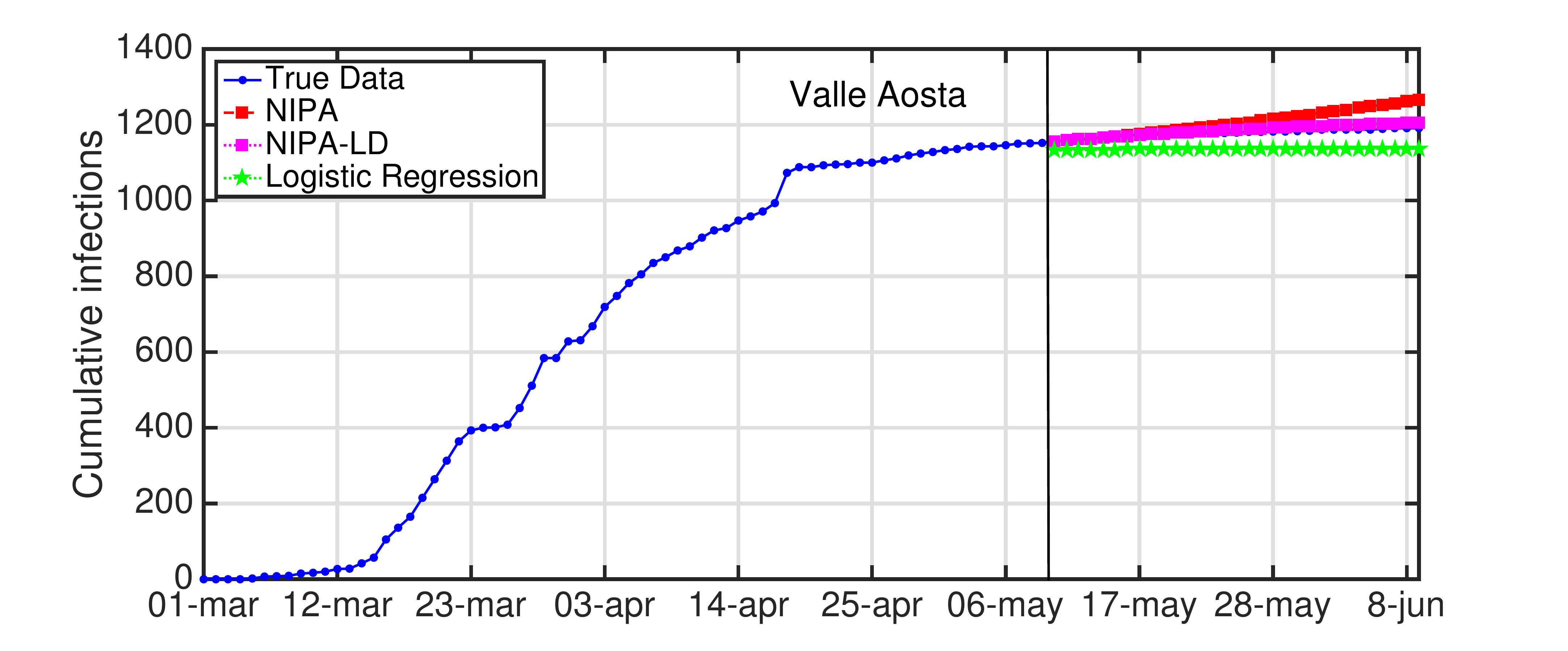}
\caption{Cumulative infections for \textit{Valle d'Aosta} with $n_{neglect}=30$.}
\label{fig:ValleA_3}
\end{figure}

\begin{figure}	
\centering
\includegraphics[width=0.7\linewidth]{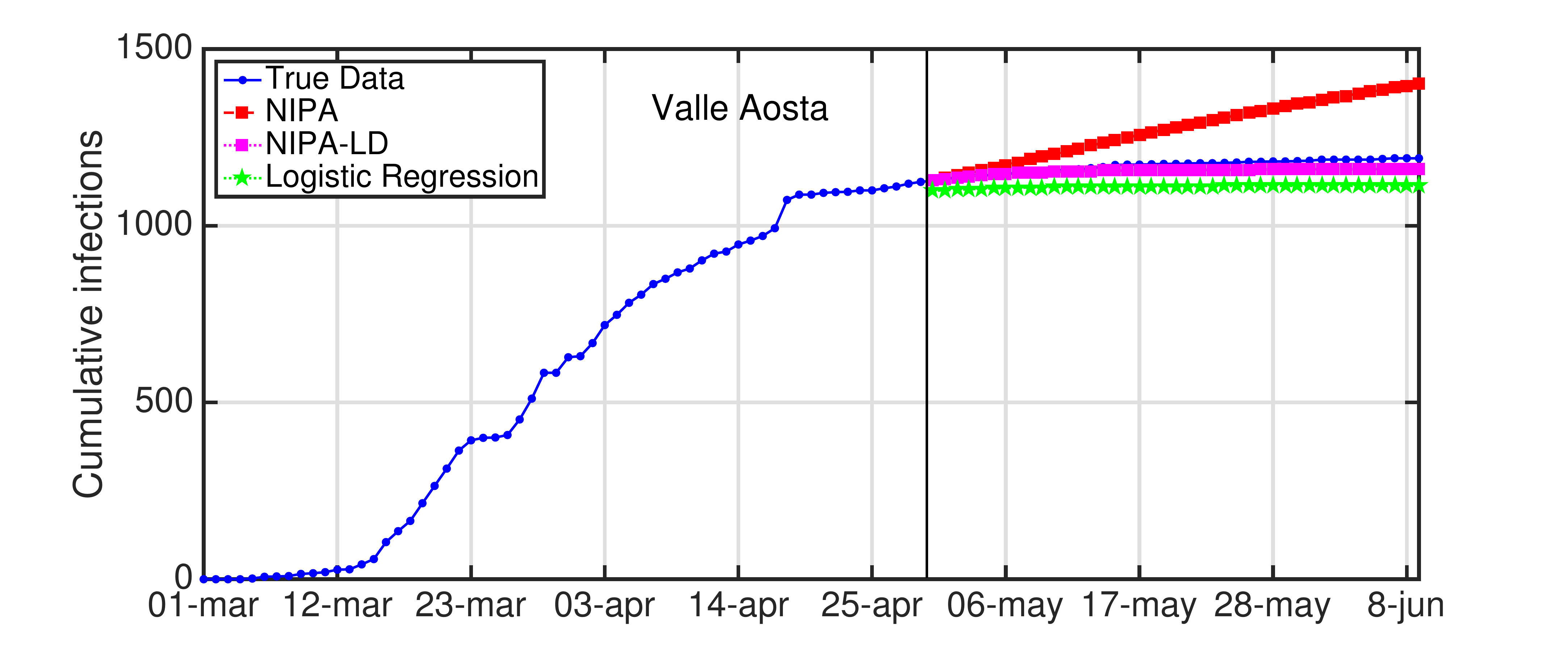}
\caption{Cumulative infections for \textit{Valle d'Aosta} with $n_{neglect}=40$.}
\label{fig:ValleA_4}
\end{figure}

\begin{figure}	
\centering
\includegraphics[width=0.7\linewidth]{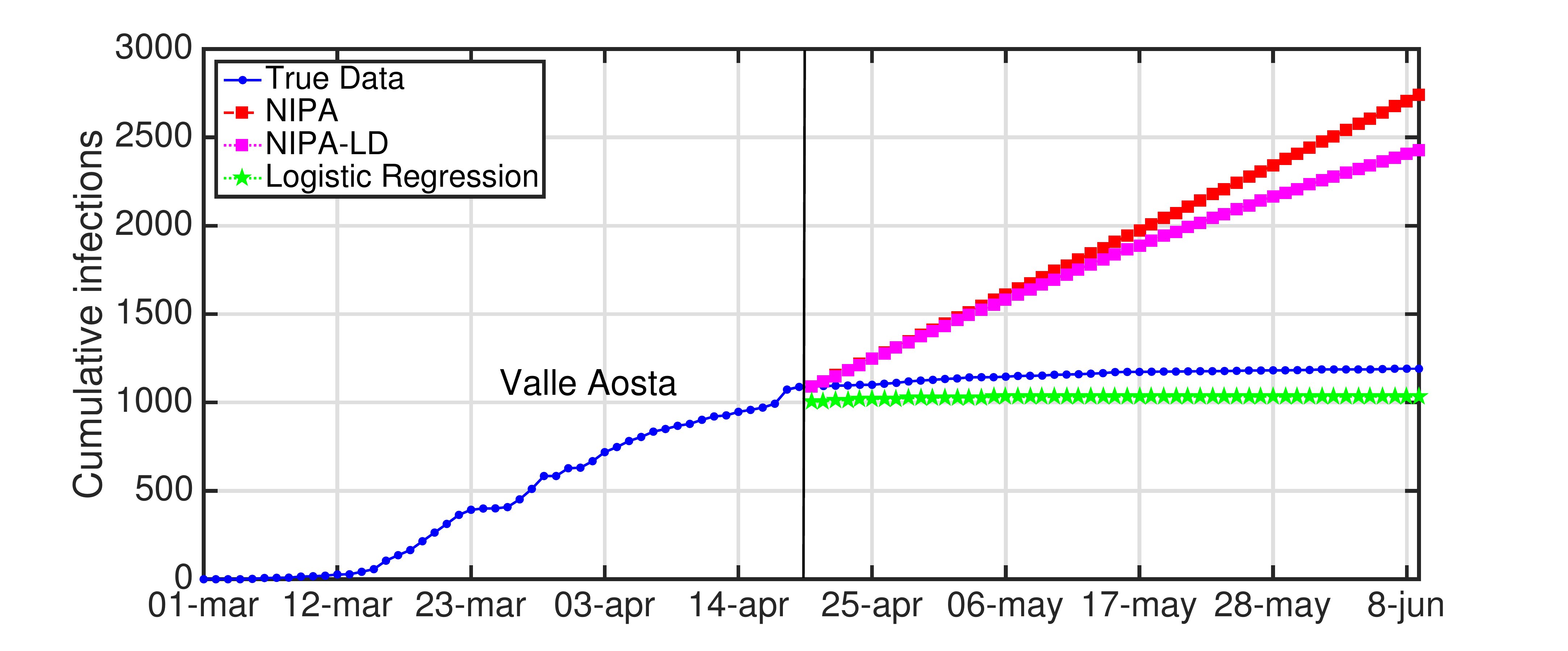}
\caption{Cumulative infections for \textit{Valle d'Aosta} with $n_{neglect}=50$.}
\label{fig:ValleA_5}
\end{figure}

Another observation is that although NIPA and NIPA-LD can obtain good short-term predictions, accurate long-term predictions are generally difficult. When aiming at predicting the infections beyond some time horizons, the accuracy of the forecasting starts decreasing. To provide a case study, in Figures \ref{fig:ValleA_1}, \ref{fig:ValleA_2}, \ref{fig:ValleA_3}, \ref{fig:ValleA_4} and \ref{fig:ValleA_5}, we show what happens when trying to predict the last 10, 20, 30, 40, 50 days of cumulative infections, respectively, in \textit{Valle d'Aosta}. In the short-term of 10 and 20 neglected days, both NIPA and NIPA-LD well match the observed data. When predicting the last 30 days until June 9, NIPA-LD predicts the infections better than NIPA. For 40 neglected days, NIPA-LD is still able to predict with a certain accuracy while NIPA definitely overestimate the cumulative infections. For 50 days, note that both the two NIPA methods are not able to accurately predict the number of cumulative infections while the logistic regression, on the contrary, works better. When thus adding too many predicted days, an accurate prediction is not possible with the NIPA-based methods. However, even if the transmission modifier is equal for all the regions, we point out that NIPA-LD performs generally better than NIPA, also for $n_{neglect}=30$ and $n_{neglect}=40$ which can be considered long-term predictions.

%\as{We finally point out that although NIPA can obtain accurate short-term predictions, it does not return a good underlying topology. For a general class of dynamics on networks (including the SIR model), completely different network topologies result in the same dynamics. Hence, it is not possible to deduce the network accurately from observations, regardless of the reconstruction method: two very different networks perfectly match the observations, and there is no reason to infer the one network instead of the other. Even though it is not possible to reconstruct the network, NIPA accurately predicts the dynamics, based on an estimated network that can be very different from the true network. This results have been shown in \cite{prasse2020predicting}, where the adjacency matrix of the network with elements $\beta_{ij}$ is assumed to be weighted, thus there is no threshold for the links.}

Finally, we point out that this work is based on the discrete-time SIR model. This model is characterized by 3 compartments. NIPA can be used for any compartmental epidemic model \cite{PrasseM19} with $c$ compartments, provided that $c-1$ compartments are measured. We \asf{underline} that the approach in this work observes only one compartment, the infectious compartment $I$, and the recovered compartment $R$ is obtained by equation (2) after estimating the curing probability $\delta_i$ in the training phase. Here, the advantage is that the less compartments we use, the less data we need to provide an accurate forecasting. When only macroscopic data, such as those exploited here, are available, a simple epidemiological model like the SIR has shown to be sufficient to predict with a high accuracy the trend of the epidemic \cite{kozyreff2020hospitalization}. More complicated  models than the SIR, such as SEIR,  SIRD, which require more additional states, do not necessarily obtain better accuracy.

\begin{table}
\centering
%\scriptsize
%\footnotesize
\tiny
\centering
\caption{Daily error fraction value between  NIPA-LD and NIPA for 30 neglected days, from day 1 to day 15.}
\label{tab:fv1p}
\begin{tabular}{|c|c|c|c|c|c|c|c|c|c|c|c|c|c|c|c|} \hline
%\hline\noalign{\smallskip}
REGION	&	1	&	2	&	3	&	4	&	5	&	6	&	7	&	8	&	9	&	10	&	11	&	12	&	13	&	14	&	15\\\hline
$Abruzzo$	&	0.95	&	0.87	&	0.60	&	0.39	&	0.29	&	1.79	&	2.36	&	3.72	&	42.05	&	12.74	&	5.05	&	3.78	&	1.69	&	1.33	&	0.89\\\hline
$Basilicata$	&	1.25	&	1.49	&	1.24	&	2.30	&	0.57	&	1.13	&	5.66	&	2.10	&	2.44	&	0.75	&	0.96	&	0.47	&	14.23	&	10.49	&	3.03\\\hline
$P.A. Bolzano$	&	0.99	&	0.99	&	0.97	&	0.96	&	0.95	&	0.94	&	0.91	&	0.90	&	0.86	&	0.85	&	0.83	&	0.81	&	0.79	&	0.76	&	0.75\\\hline
$Calabria$	&	0.29	&	1.62	&	2.25	&	2.18	&	3.88	&	1.79	&	2.41	&	3.86	&	4.09	&	3.46	&	5.59	&	7.73	&	42.64	&	11.92	&	5.20\\\hline
$Campania$	&	1.40	&	1.78	&	17.39	&	2.37	&	8.57	&	50.66	&	11.08	&	16.77	&	17.08	&	36.70	&	15.02	&	12.15	&	12.58	&	5.93	&	4.18\\\hline
$Emilia$	&	0.91	&	0.89	&	0.86	&	0.78	&	0.73	&	0.66	&	0.61	&	0.58	&	0.53	&	0.49	&	0.44	&	0.39	&	0.36	&	0.33	&	0.31\\\hline
$Friuli$	&	0.10	&	1.75	&	0.86	&	0.25	&	1.37	&	1.22	&	1.17	&	1.06	&	0.84	&	0.77	&	0.73	&	0.95	&	0.90	&	0.75	&	0.68\\\hline
$Lazio$	&	0.96	&	0.93	&	0.87	&	0.77	&	23.13	&	3.49	&	3.20	&	3.04	&	41.13	&	9.69	&	3.61	&	3.97	&	1.83	&	1.29	&	0.87\\\hline
$Liguria$	&	0.61	&	0.82	&	2.46	&	1.41	&	1.33	&	1.39	&	1.46	&	1.70	&	1.59	&	1.79	&	1.77	&	1.84	&	1.97	&	1.96	&	2.34\\\hline
$Lombardia$	&	0.96	&	1.03	&	1.08	&	1.12	&	1.26	&	1.39	&	1.64	&	2.89	&	2.64	&	3.88	&	5.70	&	11.31	&	6.56	&	11.18	&	13.02\\\hline
$Marche$	&	0.76	&	0.39	&	0.14	&	0.07	&	0.03	&	0.29	&	0.59	&	0.52	&	0.26	&	0.17	&	0.20	&	0.19	&	0.16	&	0.19	&	0.15\\\hline
$Molise$	&	1.00	&	1.00	&	1.00	&	1.00	&	1.00	&	1.00	&	0.99	&	0.99	&	0.99	&	0.99	&	0.99	&	0.99	&	0.99	&	0.99	&	0.99\\\hline
$Piemonte$	&	0.96	&	0.93	&	0.80	&	0.61	&	0.39	&	0.08	&	0.30	&	0.32	&	0.20	&	0.17	&	0.31	&	0.34	&	0.27	&	0.18	&	0.14\\\hline
$Puglia$	&	0.90	&	0.87	&	0.81	&	0.76	&	0.70	&	0.65	&	0.61	&	0.56	&	0.49	&	0.41	&	0.38	&	0.15	&	0.11	&	0.05	&	0.01\\\hline
$Sardegna$	&	0.85	&	0.84	&	0.79	&	0.75	&	0.65	&	0.50	&	0.46	&	0.45	&	0.41	&	0.38	&	0.35	&	0.34	&	0.33	&	0.33	&	0.35\\\hline
$Sicilia$	&	0.72	&	0.91	&	0.82	&	0.68	&	0.63	&	0.56	&	0.53	&	0.47	&	0.40	&	0.33	&	0.29	&	0.28	&	0.32	&	0.31	&	0.29\\\hline
$Toscana$	&	0.85	&	0.76	&	0.60	&	0.38	&	0.24	&	0.02	&	0.32	&	0.21	&	0.08	&	0.08	&	0.10	&	0.27	&	0.24	&	0.23	&	0.18\\\hline
$P.A. Trento$	&	0.88	&	0.73	&	0.45	&	0.52	&	0.52	&	0.41	&	0.22	&	0.01	&	0.02	&	0.11	&	0.20	&	0.27	&	0.27	&	0.31	&	0.21\\\hline
$Umbria$	&	0.73	&	1.28	&	3.13	&	3.07	&	1.91	&	0.43	&	0.51	&	0.23	&	0.44	&	0.25	&	0.30	&	0.20	&	0.18	&	0.12	&	0.08\\\hline
$Valle Aosta$	&	0.97	&	0.92	&	0.72	&	0.29	&	1.61	&	10.45	&	0.03	&	0.20	&	0.27	&	0.37	&	0.36	&	0.35	&	0.38	&	0.36	&	0.34\\\hline
$Veneto$	&	0.80	&	0.62	&	0.51	&	0.41	&	0.27	&	0.17	&	0.18	&	0.18	&	0.08	&	0.03	&	0.05	&	0.04	&	0.05	&	0.04	&	0.04\\\hline
\end{tabular}
\end{table}

\begin{table}
\centering
\scriptsize
%\footnotesize
\centering
\tiny
\centering
\caption{Daily error fraction value between  NIPA-LD and NIPA for 30 neglected days, from day 16 to day 30.}
\label{tab:fv2}
\begin{tabular}{|c|c|c|c|c|c|c|c|c|c|c|c|c|c|c|c|c|} \hline

%\hline\noalign{\smallskip}
REGION	&16	&	17	&	18	&	19	&	20	&	21	&	22	&	23	&	24	&	25	&	26	&	27	&	28	&	29	&	30	&	AVG\\\hline	
$Abruzzo$ &0.72	&	0.67	&	0.55	&	0.42	&	0.34	&	0.37	&	0.31	&	0.31	&	0.29	&	0.28	&	0.25	&	0.23	&	0.25	&	0.22	&	0.20	&	2.80	\\\hline
$Basilicata$&	1.66	&	1.09	&	0.78	&	0.58	&	0.45	&	0.35	&	0.28	&	0.23	&	0.18	&	0.15	&	0.12	&	0.09	&	0.07	&	0.05	&	0.09	&	1.81	\\\hline
$P.A. Bolzano$&	0.73	&	0.72	&	0.70	&	0.68	&	0.67	&	0.65	&	0.63	&	0.62	&	0.61	&	0.59	&	0.58	&	0.56	&	0.54	&	0.53	&	0.52	&	0.75	\\\hline
$Calabria$& 3.32	&	3.20	&	2.42	&	1.94	&	1.62	&	1.39	&	1.22	&	1.09	&	0.98	&	0.89	&	0.95	&	0.87	&	0.81	&	0.76	&	0.71	&	4.04	\\\hline
$Campania$ &	4.79	&	3.68	&	2.73	&	2.77	&	2.79	&	2.33	&	1.93	&	1.60	&	1.73	&	1.40	&	1.13	&	0.94	&	0.85	&	0.76	&	0.71	&	8.13	\\\hline
$Emilia$ &	0.29	&	0.29	&	0.24	&	0.21	&	0.21	&	0.19	&	0.18	&	0.18	&	0.17	&	0.17	&	0.16	&	0.16	&	0.15	&	0.15	&	0.14	&	0.39	\\\hline
$Friuli$ &	0.80	&	0.74	&	0.75	&	0.72	&	0.67	&	0.60	&	0.53	&	0.49	&	0.43	&	0.41	&	0.37	&	0.36	&	0.32	&	0.30	&	0.27	&	0.71	\\\hline
$Lazio$ &	0.71	&	0.48	&	0.49	&	0.43	&	0.29	&	0.25	&	0.21	&	0.14	&	0.12	&	0.11	&	0.09	&	0.17	&	0.16	&	0.18	&	0.23	&	3.43	\\\hline
$Liguria$ &	2.28	&	2.41	&	2.94	&	3.89	&	4.38	&	7.14	&	5.25	&	8.18	&	16.51	&	72.93	&	18.55	&	8.07	&	5.18	&	3.90	&	2.91	&	6.30	\\\hline
$Lombardia$ &	4.25	&	6.30	&	9.80	&	13.51	&	8.04	&	5.72	&	3.11	&	2.68	&	2.59	&	2.01	&	2.54	&	2.22	&	1.94	&	1.87	&	1.81	&	4.47	\\\hline
$Marche$ &	0.13	&	0.10	&	0.08	&	0.07	&	0.06	&	0.06	&	0.04	&	0.04	&	0.03	&	0.02	&	0.02	&	0.01	&	0.01	&	0.01	&	0.00	&	0.16	\\\hline
$Molise$ &	0.99	&	0.98	&	0.98	&	0.98	&	0.98	&	0.98	&	0.98	&	0.98	&	0.98	&	0.98	&	0.98	&	0.97	&	0.97	&	0.97	&	0.97	&	0.99	\\\hline
$Piemonte$ &	0.20	&	0.24	&	0.24	&	0.24	&	0.29	&	0.29	&	0.24	&	0.26	&	0.22	&	0.20	&	0.20	&	0.20	&	0.17	&	0.15	&	0.14	&	0.31	\\\hline
$Puglia$ &	0.04	&	0.01	&	0.01	&	0.03	&	0.01	&	0.01	&	0.00	&	0.02	&	0.04	&	0.03	&	0.03	&	0.02	&	0.04	&	0.04	&	0.05	&	0.26	\\\hline
$Sardegna$ &	0.34	&	0.32	&	0.31	&	0.30	&	0.29	&	0.29	&	0.27	&	0.27	&	0.27	&	0.26	&	0.24	&	0.22	&	0.22	&	0.22	&	0.22	&	0.39	\\\hline
$Sicilia$ &	0.28	&	0.25	&	0.24	&	0.24	&	0.23	&	0.23	&	0.23	&	0.21	&	0.22	&	0.22	&	0.21	&	0.21	&	0.20	&	0.20	&	0.19	&	0.36	\\\hline
$Toscana$ &	0.12	&	0.12	&	0.09	&	0.06	&	0.07	&	0.05	&	0.03	&	0.03	&	0.02	&	0.01	&	0.01	&	0.00	&	0.01	&	0.01	&	0.01	&	0.17	\\\hline
$P.A. Trento$ &	0.26	&	0.27	&	0.23	&	0.20	&	0.16	&	0.13	&	0.11	&	0.08	&	0.06	&	0.04	&	0.03	&	0.02	&	0.01	&	0.00	&	0.00	&	0.22	\\\hline
$Umbria$ &	0.08	&	0.05	&	0.02	&	0.00	&	0.01	&	0.03	&	0.04	&	0.05	&	0.06	&	0.06	&	0.07	&	0.07	&	0.07	&	0.07	&	0.05	&	0.45	\\\hline
$Valle Aosta$ &	0.30	&	0.31	&	0.30	&	0.30	&	0.29	&	0.27	&	0.23	&	0.23	&	0.24	&	0.24	&	0.24	&	0.22	&	0.20	&	0.20	&	0.20	&	0.71	\\\hline
$Veneto$ &	0.05	&	0.05	&	0.05	&	0.05	&	0.05	&	0.05	&	0.06	&	0.06	&	0.07	&	0.07	&	0.07	&	0.07	&	0.07	&	0.08	&	0.08	&	0.15	\\\hline
\end{tabular}
\end{table}

\section*{Conclusion} \label{sec:concl}
We exploited a network-based SIR model to predict the curves of the cumulative infections of individuals affected by the SARS-CoV-2 virus in Italy. The classic SIR epidemic model has been expanded by incorporating time-varying lockdown protocols in order to have epidemic transmission rates that change as the government quarantine rules change. 
Tested on regional data of the COVID-19 in Italy, the network-based prediction method results in a higher prediction accuracy when compared to the classical method that does not consider the lockdown measures. 

Experiments, however,  pointed out that equal values of the transmission modifiers for all the Italian regions could not be appropriate, because of the differences in people mobility.
On the other hand, the NIPA method extended to account for the  lockdown measures highlighted the tremendous potential of an optimal transmission modifier. In fact NIPA-LD  could be practically used to experiment which lockdown strategies are effective or not and which countermeasures are more appropriate to stop the spreading of COVID-19 epidemic.
Future work will investigate  how a transmission modifier might be best related to a quarantine strategy also in the training phase of NIPA, in order to improve the prediction capability of the approach.

\section*{Acknowledgements}
This work has been supported by the Universiteits fonds Delft under the program  TU Delft Covid-19 Response Fund.

\bibliographystyle{plain}

\bibliography{main}

\begin{thebibliography}{10}

\bibitem{Arenas2020}
A.~Arenas, W.~Cota, J.~Gomez-Gardenes, S.~G\'omez, C.~Granell, J.~T. Matamalas,
  D.~Soriano-Panosand, and B.~Steinegger.
\newblock A mathematical model for the spatiotemporal epidemic spreading of
  {COVID19}.
\newblock {\em medRxiv:10.1101/2020.03.21.20040022v1.}, March 2020.

\bibitem{caccavo2020chinese}
Diego Caccavo.
\newblock Chinese and italian covid-19 outbreaks can be correctly described by
  a modified sird model.
\newblock {\em medRxiv}, 2020.

\bibitem{Chinazzi2020}
M.~Chinazzi, J.~T. Davis, M.~Ajelli, C.~Gioannini, M.~Litvinova, S.~Merler,
  A.~Pastore y~Piontti, K.~Mu, L.~Rossi, K.~Sun, C.~Viboud, X.~Xiong, H.~Yu,
  M.~E. Halloran, I.~M.~Longini Jr., and A.~Vespignani.
\newblock The effect of travel restrictions on the spread of the 2019 novel
  coronavirus ({COVID}-19) outbreak.
\newblock {\em Science}, 368(6489):395--400, 2020.

\bibitem{Chu_Lancet2020}
D.~K. Chu, E.~A. Akl, S.~Duda, K.~Solo, S.~Yaacoub, and H.~J. Sch{\"u}nemann.
\newblock Physical distancing, face masks, and eye protection to prevent
  person-to-person transmission of {SARS-CoV-2} and {COVID-19}: a systematic
  review and meta-analysis.
\newblock {\em The Lancet}, pages doi.org/10.1016/S0140--6736(20)31142--9, June
  1, Published online 2020.

\bibitem{di2020impact}
L~Domenico Di, G~Pullano, CE~Sabbatini, PY~Bo{\"e}lle, and V~Colizza.
\newblock Impact of lockdown on covid-19 epidemic in {\^i}le-de-france and
  possible exit strategies.
\newblock {\em BMC medicine}, 18(1):240--240, 2020.

\bibitem{estrada2020covid}
Ernesto Estrada.
\newblock Covid-19 and sars-cov-2. modeling the present, looking at the future.
\newblock {\em Physics Reports}, 2020.

\bibitem{ferrari2020modelling}
Luisa Ferrari, Giuseppe Gerardi, Giancarlo Manzi, Alessandra Micheletti,
  Federica Nicolussi, and Silvia Salini.
\newblock Modelling provincial covid-19 epidemic data in italy using an
  adjusted time-dependent sird model.
\newblock {\em arXiv preprint arXiv:2005.12170}, 2020.

\bibitem{flaxman2020estimating}
Seth Flaxman, Swapnil Mishra, Axel Gandy, H~Juliette~T Unwin, Thomas~A Mellan,
  Helen Coupland, Charles Whittaker, Harrison Zhu, Tresnia Berah, Jeffrey~W
  Eaton, et~al.
\newblock Estimating the effects of non-pharmaceutical interventions on
  covid-19 in europe.
\newblock {\em Nature}, 584(7820):257--261, 2020.

\bibitem{Galeazzi2020}
A.~Galeazzi, M.~Cinelli, G.~Bonaccorsi, F.~Pierri, A.~L. Schmidt, A.~Scala,
  F.~Pammolli, and W.~Quattrociocchi.
\newblock Human mobility in response to {COVID}-19 in france, italy and uk.
\newblock {\em arXiv preprint arXiv:2005.06341}, May 2020.

\bibitem{giuliani2020modelling}
Diego Giuliani, Maria~Michela Dickson, Giuseppe Espa, and Flavio Santi.
\newblock Modelling and predicting the spatio-temporal spread of coronavirus
  disease 2019 (covid-19) in italy.
\newblock {\em Available at SSRN 3559569}, 2020.

\bibitem{hadjidemetriou2020impact}
Georgios~M Hadjidemetriou, Manu Sasidharan, Georgia Kouyialis, and Ajith~K
  Parlikad.
\newblock The impact of government measures and human mobility trend on
  covid-19 related deaths in the uk.
\newblock {\em Transportation research interdisciplinary perspectives}, page
  100167, 2020.

\bibitem{haug2020ranking}
Nils Haug, Lukas Geyrhofer, Alessandro Londei, Elma Dervic, Amelie
  Desvars-Larrive, Vittorio Loreto, Beate Pinior, Stefan Thurner, and Peter
  Klimek.
\newblock Ranking the effectiveness of worldwide covid-19 government
  interventions.
\newblock {\em MedRxiv}, 2020.

\bibitem{Hethcote00}
H.~W. Hethcote.
\newblock The mathematics of infectious diseases.
\newblock {\em {SIAM} Review}, 42(4):599--653, 2000.

\bibitem{Kermack}
W.~O. Kermack and A.G. A.~G.~McKendrick.
\newblock A contribution to the mathematical theory of epidemics.
\newblock In {\em Proceedings of the Royal Society of London. Series A}, volume
  115, page 700?721, 1927.

\bibitem{Klein2020}
B.~Klein, T.~LaRock, S.~McCabe, L.~Torres, F.~Privitera, B.~Lake, M.~U.~G.
  Kraemer, J.~S. Brownstein, D.~Lazer, T.~Eliassi-Rad, S.~V. Scarpino,
  M.~Chinazzi, and A.~Vespignani.
\newblock Assessing changes in commuting and individual mobility in major
  metropolitan areas in the united states during the {COVID}-19 outbreak
  (2020).
\newblock {\em preprint on webpage at
  https://www.networkscienceinstitute.org/publications}, 2020.

\bibitem{kozyreff2020hospitalization}
Gregory Kozyreff.
\newblock Hospitalization dynamics during the first covid-19 pandemic wave: Sir
  modelling compared to belgium, france, italy, switzerland and new york city
  data.
\newblock {\em arXiv preprint arXiv:2007.01411}, 2020.

\bibitem{lavezzo2020suppression}
Enrico Lavezzo, Elisa Franchin, Constanze Ciavarella, Gina Cuomo-Dannenburg,
  Luisa Barzon, Claudia Del~Vecchio, Lucia Rossi, Riccardo Manganelli, Arianna
  Loregian, Nicol{\`o} Navarin, et~al.
\newblock Suppression of covid-19 outbreak in the municipality of vo, italy.
\newblock {\em medRxiv}, 2020.

\bibitem{MaierScience2020}
B.~F. Maier and D.~Brockmann.
\newblock Effective containment explains subexponential growth in recent
  confirmed {COVID}-19 cases in {C}hina.
\newblock {\em Science}, 4557(eabb4557), 2020.

\bibitem{Oliver2020}
N.~Oliver, B.~Lepri, H.~Sterly, R.~Lambiotte, S.~Deletaille, M.~De Nadai,
  E.~Letouze, A.~A. Salah, R.~Benjamins, C.~Cattuto, V.~Colizza, N.~de~Cordes,
  S.~P. Fraiberger, T.~Koebe, S.~Lehmann, J.~Murillo, A.~Pentland, P.~N. Pham,
  F.~Pivetta, J.~Saramaki, S.~V. Scarpino, M.~Tizzoni, S.~Verhulst, and
  P.~Vinck.
\newblock Mobile phone data for informing public health actions across the
  {COVID}-19 pandemic life cycle.
\newblock {\em Science Advances}, 6(23), 2020.

\bibitem{PVM_RMP_epidemics2014}
R.~Pastor-Satorras, C.~Castellano, P.~Van~Mieghem, and A.~Vespignani.
\newblock Epidemic processes in complex networks.
\newblock {\em Review of Modern Physics}, 87(3):925--979, September 2015.

\bibitem{pei2020differential}
Sen Pei, Sasikiran Kandula, and Jeffrey Shaman.
\newblock Differential effects of intervention timing on covid-19 spread in the
  united states.
\newblock {\em medRxiv}, 2020.

\bibitem{PrasseM19}
B.~Prasse and P.~Van~Mieghem.
\newblock Network reconstruction and prediction of epidemic outbreaks for
  general group-based compartmental epidemic models.
\newblock {\em IEEE Transactions on Network Science and Engineering, to
  appear}, 2020.

\bibitem{Prasse2020}
Bastian Prasse, Massimo~A. Achterberg, Long Ma, and Piet Van~Mieghem.
\newblock Network-inference-based prediction of the {COVID}-19 epidemic
  outbreak in the {C}hinese province {H}ubei.
\newblock {\em Applied Network Science}, 5(1):1--11, 2020.

\bibitem{prasse2020fundamental}
Bastian Prasse and P.~Van~Mieghem.
\newblock Fundamental limits of predicting epidemic outbreaks.
\newblock {\em Delft University of Technology}, 2020.

\bibitem{prasse2020predicting}
Bastian Prasse and Piet Van~Mieghem.
\newblock Predicting dynamics on networks hardly depends on the topology.
\newblock {\em arXiv preprint arXiv:2005.14575}, 2020.

\bibitem{Schlosser2020}
F.~Schlosser, B.~F. Maier, D.~Hinrichs, A.~Zachariae, and D.~Brockmann.
\newblock {COVID}-19 lockdown induces structural changes in mobility networks?
  {I}mplication for mitigating disease dynamics.
\newblock {\em arXiv preprint arXiv:2007.01583v2}, July 2020.

\bibitem{Wang2020}
P.~X. Song, L.~Wang, Y.~Zhou, J.~He, B.~Zhum, F.~Wang, L.~Tang, and
  M.~Eisenberg.
\newblock An epidemiological forecast model and software assessing
  interventions on {COVID}-19 epidemic in {C}hina.
\newblock {\em med{R}xiv, preprint doi:
  https://doi.org/10.1101/2020.02.29.20029421, march 2020}, 2020.

\bibitem{Scoglio2011}
M.~Youssef and C.~Scoglio.
\newblock An individual-based approach to {SIR} epidemics in contact networks.
\newblock {\em Journal of Theoretical Biology}, 283(1):136--144, 2011.

\end{thebibliography}

\end{document}